%% file: main.tex
\documentclass[a4paper,dvipsnames,11pt]{article}
\input{topmatter}
\input{titlepage}

\begin{document}
\maketitle
\flushbottom

\input{sections/sec_1_intro}
\input{sections/sec_2_ADEalgebras}
\input{sections/sec_3_graded_unitarity.tex}

\input{sections/sec_4_typeA_graded_unitarity}
\input{sections/sec_5_typeD_graded_unitarity}
\input{sections/sec_6_exceptional_gr_unit}
%\input{sections/sec_7_outlook}

%%%%%%%%%%%%%%%%%%%%%%%%%%%%%%%%%%%%%%%%%%%%%%%%%%%%%%%%%%%%%%%%%%%%%%%%

\acknowledgments

The author thanks Tomoyuki Arakawa and Palash Singh for helpful discussions related to this work, and is particularly grateful to Christopher Beem and Niklas Garner for discussions at various stages of the project and insightful comments on the manuscript. The author acknowledges the use of \texttt{GPT-5.6 Sol} for fixing typographical errors and identifying relevant references; in particular it brought the reference \cite{panyushev2007semi} to the author's attention. All ideas, arguments, and computations in the paper are the author's own. This work was supported by the Clarendon Fund Scholarship and the Eddie Dinshaw Scholarship at Balliol College.

%%%%%%%%%%%%%%%%%%%%%%%%%%%%%%%%%%%%%%%%%%%%%%%%%%%%%%%%%%%%%%%%%%%%%%%%

\vfill\eject

\appendix
\input{sections/appendix_VkSOodd}

%%%%%%%%%%%%%%%%%%%%%%%%%%%%%%%%%%%%%%%%%%%%%%%%%%%%%%%%%%%%%%%%%%%%%%%%
\bibliographystyle{aux/JHEP}
\bibliography{aux/references}
%%%%%%%%%%%%%%%%%%%%%%%%%%%%%%%%%%%%%%%%%%%%%%%%%%%%%%%%%%%%%%%%%%%%%%%%

\end{document}

%% file: titlepage.tex
\title{Graded unitarity constraints on simply-laced affine Kac--Moody algebras}
\author{Harshal Kulkarni}
\affiliation{Mathematical Institute, University of Oxford, Woodstock Road, Oxford, OX2 6GG, UK}
\abstract{We investigate constraints imposed by four-dimensional unitarity (recently formalised as graded unitarity) on possible simply-laced affine Kac--Moody vertex algebras arising from four dimensions via the SCFT/VOA correspondence. Under the assumption that the $\mathfrak{R}$-filtration is weight-based with respect to the currents and the Casimir operators, with appropriately shifted $R$-charge assignments for the latter, we demonstrate that all values of the level other than a discrete set are incompatible with four-dimensional unitarity. This set includes all the boundary admissible affine vertex algebras, those corresponding to rank-one Kodaira SCFTs, as well as certain non-admissible families conjecturally associated with specific $D^b_q(\gf)$ Argyres--Douglas theories. Specialising to $\mathfrak{sl}_N$ with $N$ prime, we find a stronger result: under the same $\mfR$-filtration assumption, four-dimensional unitarity rules out all levels except the boundary admissible $k = k_{N,q}$, which are precisely those corresponding to the $D_q^N(\mathfrak{sl}_N)$ Argyres--Douglas theories.}

%% file: sections/sec_1_intro.tex
%!TEX Root = ../main.tex

%%%%%%%%
\section{\label{sec:intro}Introduction and Summary}
%%%%%%%%
Every four-dimensional $\mathcal{N} = 2$ superconformal field theory (SCFT) gives rise to an associated vertex operator algebra (VOA), via the SCFT/VOA correspondence \cite{Beem:2013sza}. These vertex algebras possess a number of nice properties that make them better behaved than generic VOAs. For example, they are conjecturally quasi-lisse so their vacuum characters satisfy finite-order modular linear differential equations \cite{Beem:2017ooy,Arakawa:2016hkg} and in many cases they admit extremely nice free field realisations \cite{Beem:2019tfp,Beem:2019snk,Beem:2021jnm,Beem:2024fom}. A salient feature is that when the four-dimensional theory in question is unitary, the associated VOA inherits a novel form of unitarity, christened \emph{graded unitarity} in \cite{ArabiArdehali:2025fad}. The implications of this structure for Virasoro, $\mathfrak{sl}_2, \mathfrak{sl}_3, \mathfrak{sl}_4$ affine vertex algebras were also investigated \emph{loc. cit.}, while the more technically involved case of Zamolodchikov $\mathcal{W}_3$ algebras was examined in \cite{Beem:2026lkq}. The variations of this structure, along with a more mathematically precise formulation were elaborated in \cite{Beem:2025guj}. Some preliminary consequences of four-dimensional unitarity for the associated vertex operator algebra were studied in \cite{Liendo:2015ofa,Beem:2018duj}, which in retrospect are consequences of graded unitarity. More recently, it was established that graded unitarity of $\mathfrak{sl}_2$ at boundary admissible level $k = -2 + \frac{2}{q}$ implies graded unitarity of the $(2,q)$ Virasoro minimal models \cite{Popov:2026qxn}. 

A graded unitarity structure for a vertex algebra $\mathbb{V}$ amounts, in particular, to a conjugation operation $\rho$ (generally of order four) together with a nondegenerate half-integral good filtration $\mathfrak{R}_{\bullet}$ on $\mathbb{V}$. With respect to these ingredients, graded unitarity amounts to the requirement that a suitably defined Hermitian form is positive definite; equivalently, that the constants appearing in the two-point functions of operators with their conjugates have definite, $R$-charge-dependent signs. It is an important problem to characterise to what extent the existence of a graded unitary structure constrains the underlying vertex algebra. A direct attack on the problem (for a fixed underlying vertex algebra) is complicated by the dependence of the positivity property on the choice of $\mathfrak{R}$-filtration. At present, there is no intrinsic characterisation of the physically relevant $\mathfrak{R}$-filtration without reference to four-dimensional input. One is thus led to an interesting (but seemingly difficult) problem of constraining a choice of good filtrations subject to highly specific sign requirements for two-point functions.

If, however, an $\mathfrak{R}$-filtration is given or assumed, then graded unitarity makes extremely precise predictions about the positivity/negativity conditions for two-point functions in the underlying vertex algebra, and these are comparatively accessible. This approach was pursued in \cite{ArabiArdehali:2025fad, Beem:2026lkq} for the cases where the underlying vertex algebra is Virasoro, an affine Kac--Moody VOA of type $\mathfrak{sl}_n$ with $n=2,3,4$, or the Zamolodchikov $\mathcal{W}_3$ VOA. In these works, a simple but physically motivated choice of $\mfR$-filtration was assumed and consequently all but a discrete set of values for the central charge/level were ruled out by imposing the relevant positivity bounds level by level. These stringent restrictions on the central charge/level were obtained by examining the positivity constraints visible at the level of the Kac determinant at a given conformal level (and $\mathfrak{g}$-weight in the Kac--Moody case). While the full set of positivity constraints from graded unitarity are much richer, it is particularly satisfying that determinant-based constraints provide such strict restrictions.

The purpose of the present paper is to generalise such a classification to \emph{all} simply-laced affine Kac--Moody algebras. There are several families of four-dimensional theories which are known, or conjectured, to give rise to simply-laced affine Kac--Moody algebras. In particular, the VOAs associated with the Kodaira/Deligne--Cvitanovi\'{c} rank-one SCFTs \cite{Seiberg:1994aj, Argyres:1995xn, Minahan:1996fg, Minahan:1996cj} are given by affine Kac--Moody VOAs $V_k(\gf)$ at level $k = -\frac{h^\vee_\gf}{6}-1$, where $h^\vee$ is the dual Coxeter number of the corresponding simple Lie algebra $\gf$. For simply-laced $\gf$, these include the family,
\begin{equation}
    \mathfrak{a}_1 \subset \mathfrak{a}_2 \subset \mathfrak{d}_4 \subset \mathfrak{e}_6 \subset \mathfrak{e}_7 \subset \mathfrak{e}_8~,
\end{equation}
so the respective affine Kac--Moody VOAs are expected to be graded unitary. 

Furthermore, a plethora of generalised Argyres--Douglas (AD) theories \cite{Argyres:1995jj, Argyres:1995xn, Xie:2012hs} conjecturally give rise to affine Kac--Moody VOAs. Specifically, the theories of interest are associated to irregular Hitchin systems of type $ADE$ with a full regular puncture and an irregular puncture \cite{Wang:2015mra}. We denote these theories by $D^b_q(\gf)$, where $b$ and $q$ parametrise the irregular singularity, and $\gf$ is simply laced.\footnote{Here, we use a different parametrisation from the one used in \cite{Wang:2015mra,Xie:2016evu, Song:2017oew}, where the theories are parametrised by $k$ and $b$. The two conventions are related by $k = -b+q$.} Largely based on the requirement that the four-dimensional theory has precisely $\gf$-flavour symmetry, and the satisfaction of the Sugawara relation for the central charge, the associated VOAs for the $D_q^{b}(\gf)$ generalised Argyres--Douglas SCFTs without extra mass parameters/flavour symmetries from the irregular puncture, are conjectured to be the affine Kac--Moody VOAs \cite{Xie:2016evu, Song:2017oew, Xie:2017aqx},
\begin{equation} \label{eq:DqbgVOA}
    \mathbb{V}[D_q^{b}(\gf)] = V_{k}(\gf), \qquad k = -h^\vee + \frac{b}{q}~. 
\end{equation}
It is worth emphasising that (non-trivial) finite extensions by modules of affine Kac--Moody algebras do \emph{not} affect the central charge $c_{4d}$ of the putative four-dimensional theory. Consequently, the agreement of the central charge alone does not distinguish the conjectured identification in \eqref{eq:DqbgVOA} from possible finite extensions. However, there is a particularly important specialisation: $b = h^\vee$, with $(h^\vee, q) = 1$, in which case the level $k$ is \emph{boundary admissible}.\footnote{For a simply-laced Lie algebra, the level $k$ is said to be admissible if it is of the form $k = -h^\vee + \frac{p}{q}$, where $p,q \in \mathbb{Z}_{\geqslant 1}$, $p \geqslant h^\vee$ and $(p,q) = 1$. When additionally, $p = h^\vee$, the level is said to be boundary admissible.} In addition to the absence of mass parameters, the four-dimensional $D_q^{h^\vee}(\gf)$ theories in this case have no exactly marginal couplings. With the stress tensor of the VOA $\mathbb{V}[D_q^{h^\vee}(\gf)]$ being the Segal--Sugawara stress tensor of $V_{k}(\gf)$ \cite{Beem:2018duj}, certain technical niceness assumptions imply that the affine Kac--Moody subalgebra $V_{k}(\gf)$ does not admit non-trivial finite extensions \cite{Arakawa:2021ogm}. Therefore, it is also expected that all the boundary admissible affine Kac--Moody VOAs appearing in \eqref{eq:DqbgVOA} are graded unitary. Indeed, boundary admissibility has already been observed to be a recurring feature of affine Kac--Moody VOAs arising from four-dimensional unitary theories \cite{ArabiArdehali:2025fad, Beem:2026lkq}. 

Since deriving the $\mathfrak{R}$-filtration from first principles is beyond the scope of this work, we make the working assumption that the $\mathfrak{R}$-filtration on the affine Kac--Moody VOA $V_k(\gf)$ is weight-based, with appropriate adjustments for Casimir operators, as introduced in \cite{ArabiArdehali:2025fad}. It is worth emphasising that affine Kac--Moody VOAs with different, \emph{non-standard} $\mfR$-filtrations \emph{do} arise in four-dimensional unitary theories \cite{Beem:2026peg}. In all known examples, however, the putative four-dimensional theories are discrete gaugings of a parent unitary theory. In \cite{ArabiArdehali:2025fad}, it was further conjectured that the standard $\mathfrak{R}$-filtration is \emph{good}; a first contribution of this paper is to prove this conjecture for \emph{any} semisimple $\gf$. 

We then restrict to simply-laced $\gf$ and examine constraints on the level $k$ from graded unitarity. In line with \cite{ArabiArdehali:2025fad}, we utilise the closed-form expression for the vacuum determinant of affine Kac--Moody algebras \cite{gorelik2007simplicity}, and take as our diagnostic the determinant-level consequence of graded unitarity. At each conformal level $L$ (and a specific fixed $\gf$-weight), we explicitly determine the relevant structure of zeros and multiplicities of the vacuum determinant. Graded unitarity predicts a definite sign for the determinant of the Shapovalov form, determined by the $R$-charges (after refining the $\mfR$-filtration to a grading). We can compare this sign prediction with the sign implied by the closed-form expression of the vacuum determinant. 

%%%%%%%%%%%%%%%%%%%%%%%%%%%%%%%%%%%%%
\input{classificationtable}
%%%%%%%%%%%%%%%%%%%%%%%%%%%%%%%%%%%%%

Our main result is that this comparison leads to extremely stringent restrictions on the level $k$: under the standard $\mfR$-filtration hypothesis, we find that all levels except the discrete set of values tabulated in Table \ref{tab:classification} are incompatible with graded unitarity. This set includes the affine Kac--Moody algebras associated with the rank-one Kodaira SCFTs and the ``class I" $D_q^{h^\vee}(\gf)$ generalised Argyres--Douglas theories with simply-laced flavour symmetry. Interestingly, determinant-based graded unitarity constraints in a specific fixed $\gf$-weight sector also admit certain \emph{non-admissible} affine Kac--Moody VOAs conjecturally associated with specific $D_q^b(\gf)$ Argyres--Douglas theories. Although finer state-based graded unitarity constraints may rule out some of these levels, their survival provides suggestive evidence for the respective conjectured identifications in \eqref{eq:DqbgVOA}. Notably, all the $D_q^b(\gf)$ Argyres--Douglas theories making an appearance in Table \ref{tab:classification} have no additional mass parameters/flavour symmetries arising from the irregular puncture.

More broadly, the present results reinforce an emerging picture that four-dimensional unitarity serves as a remarkably powerful rigidity principle, sharply constraining the vertex algebras appearing in the image of the SCFT/VOA map. In all the examples studied so far, Virasoro, $\mathcal{W}_3$, and now all simply-laced affine VOAs, even the determinant-based consequences of graded unitarity already single out a limited, discrete set of values that can possibly arise from four-dimensional theories. We expect similar classifying results to hold for the non-simply laced affine VOAs as well. These results, while capturing a coarse shadow of the full positivity constraints encoded by graded unitarity, point towards the need for an intrinsic vertex-algebraic formulation of graded unitarity. A central question is to determine which structural properties of a vertex algebra govern the existence of a compatible $\mfR$-filtration and conjugation satisfying the appropriate positivity conditions. 

The organisation of the paper is as follows. In Section \ref{sec:GKvacdet} we review the relevant aspects of the Gorelik--Kac vacuum determinant for simply-laced affine Kac--Moody algebras, and reformulate the functions encoding the multiplicities of zeros in a form better suited for explicit computations. In Section \ref{sec:grunitstructures} we specialise the general framework of graded unitarity to affine Kac--Moody VOAs, describe the standard $\mfR$-filtration, and prove its goodness. We also state the determinant-based sign constraint in the simply laced case. In Section \ref{sec:typeAgrunit} we analyse graded unitarity constraints for $V_k(\mathfrak{sl}_N)$ affine Kac--Moody VOAs. We begin with a bespoke low-level analysis, and then study the structure of zeros of the vacuum determinant to deduce the constraints on the level $k$ and obtain the aforementioned restriction. Section \ref{sec:typeDgrunit} carries out the analogous analysis for $V_k(\mathfrak{so}_{2N})$. After treating $V_k(\mathfrak{so}_8)$ explicitly, we analyse the case of even $N$, and then summarise the complete classification, with the details for odd $N$ relegated to Appendix \ref{app:VkSOodd}. Finally, in Section \ref{sec:typeEgrunit} we complete the characterisation of graded unitarity constraints for simply-laced affine Kac--Moody VOAs by determining the determinant-based constraints for $V_k(\mathfrak{e}_{n=6,7,8})$. 

%% file: classificationtable.tex
\begin{table}[]
\centering
\footnotesize

\renewcommand{\arraystretch}{1.60}
\setlength{\tabcolsep}{4pt}

\setlength{\arrayrulewidth}{0.6pt}
\arrayrulecolor{black}

%\begin{tabularx}
\begin{tabularx}{\textwidth}{
|>{\centering\arraybackslash}p{0.16\textwidth}
|>{\raggedright\arraybackslash}p{0.22\textwidth}
|>{\raggedright\arraybackslash}p{0.33\textwidth}
|X|
}
\hline

\textbf{Vertex algebra} & \centering \textbf{Level $k$} & \centering \textbf{Conditions} & \textbf{SCFT realisation} \\
\hline

% A
\multirow{2}{*}{$V_k(\mathfrak{sl}_N)$} & \cellcolor{knownSCFT}$-N+\frac{N}{q}$ & \cellcolor{knownSCFT}$(N,q)=1$ & \cellcolor{knownSCFT}``class I'' $D_q^N(\mathfrak{sl}_N)$ \\
\hhline{~---}

& $-N+\frac{N}{q}$ & $(N,q)>1,\quad N\nmid q$ & \\
\hline

% D
\multirow{8}{*}{\shortstack{$V_k(\mathfrak{so}_{2N})$ \\ $(N\geqslant4)$
}} & \cellcolor{knownSCFT}$-(2N-2)+\frac{2N-2}{q}$ & \cellcolor{knownSCFT}$(2N-2,q)=1$ & \cellcolor{knownSCFT}``class I'' $D_q^{2N-2}(\mathfrak{so}_{2N})$ \\
\hhline{~---}

& \cellcolor{newSCFT} $-(2N-2)+\frac{2N-2}{q}$ & \cellcolor{newSCFT} $q\in2\mathbb N+1,\quad(2N-2,q)>1$ & \cellcolor{newSCFT} \qquad $D_q^{2N-2}(\mathfrak{so}_{2N})$ \\
\hhline{~---}

& \cellcolor{knownSCFT}$-2$ & \cellcolor{knownSCFT}$N=4$ & \cellcolor{knownSCFT}$SU(2)$ with $N_f=4$ \\
\hhline{~---}

& \cellcolor{newSCFT} $-(2N-2)+\frac{N}{q}$ & \cellcolor{newSCFT} $1\leqslant q\leqslant \frac{N-2}{2}, \ \ \frac{N}{\text{gcd}(N,q)} \in 2\mathbb{N}$ & \cellcolor{newSCFT} \qquad $D_q^{N}(\mathfrak{so}_{2N})$ \\
\hhline{~---}

& $-(2N-2)+\frac{N}{q}$ & $1\leqslant q\leqslant \frac{N-2}{2}, \ \ \frac{N}{\text{gcd}(N,q)} \in 2\mathbb{N}+1$ &  \\
\hhline{~---}

& \cellcolor{newSCFT} $-(2N-2)+\frac{2N}{3N-2}$ & \cellcolor{newSCFT} $N\in4\mathbb N$ & \cellcolor{newSCFT} \qquad $D_{\frac{3N-2}{2}}^{N}(\mathfrak{so}_{2N})$ \\
\hhline{~---}

& $-(2N-2)+\frac{2N}{3N-2}$ & $N\in4\mathbb N+2$ & \\
\hhline{~---}

& $-(2N-2)+\frac{2}{q}$ & $q\in2\mathbb N+1$, \quad $N \in 2\mathbb{N}+1$ & \\
\hhline{~---}

& $-\frac{5}{2}$ & $N=4$ & \\
\hhline{~---}

& $-\frac{22}{3}$ & $N=6$ & \\
\hline

% E6
\multirow{4}{*}{$V_k(\mathfrak e_6)$} & \cellcolor{knownSCFT}$-12+\frac{12}{q}$ & \cellcolor{knownSCFT}$(12,q)=1$ & \cellcolor{knownSCFT}``class I'' $D_q^{12}(\mathfrak e_6)$ \\
\hhline{~---}

&  \cellcolor{newSCFT} $-12+\frac{3}{q}$ & \cellcolor{newSCFT} $(3,q)=1$ &  \cellcolor{newSCFT} \qquad $D_{4q}^{12}(\mathfrak e_6)$ \\
\hhline{~---}

& \cellcolor{knownSCFT}$-3$ & \cellcolor{knownSCFT} & \cellcolor{knownSCFT}rank-one $E_6$ SCFT \\
\hhline{~---}

& \cellcolor{newSCFT} $-6$ & \cellcolor{newSCFT} & \cellcolor{newSCFT} \qquad $D_{2}^{12}(\mathfrak e_6)$ \\
\hline

% E7
\multirow{5}{*}{$V_k(\mathfrak e_7)$} & \cellcolor{knownSCFT}$-18+\frac{18}{q}$ & \cellcolor{knownSCFT}$(18,q)=1$ & \cellcolor{knownSCFT}``class I'' $D_q^{18}(\mathfrak e_7)$ \\
\hhline{~---}

& \cellcolor{newSCFT} $-18+\frac{2}{q}$ & \cellcolor{newSCFT} $(2,q)=1$ & \cellcolor{newSCFT} \qquad $D^{18}_{9q}(\mathfrak{e}_7)$ \\
\hhline{~---}

& \cellcolor{knownSCFT}$-4$ & \cellcolor{knownSCFT} & \cellcolor{knownSCFT}rank-one $E_7$ SCFT \\
\hhline{~---}

& $-8$ & & \\
\hhline{~---}

& $-11$ & & \\
\hline

% E8
\multirow{6}{*}{$V_k(\mathfrak e_8)$} & \cellcolor{knownSCFT}$-30+\frac{30}{q}$ & \cellcolor{knownSCFT}$(30,q)=1$ & \cellcolor{knownSCFT}``class I'' $D_q^{30}(\mathfrak e_8)$ \\
\hhline{~---}

& \cellcolor{knownSCFT}$-6$ & \cellcolor{knownSCFT} & \cellcolor{knownSCFT}rank-one $E_8$ SCFT \\
\hhline{~---}

& \cellcolor{newSCFT} $-10$ & \cellcolor{newSCFT} & \cellcolor{newSCFT} \qquad $D_1^{20}(\mathfrak e_8)$ \\
\hhline{~---}

& \cellcolor{newSCFT} $-18$ & \cellcolor{newSCFT} & \cellcolor{newSCFT} \qquad $D_2^{24}(\mathfrak e_8)$ \\
\hhline{~---}

& \cellcolor{newSCFT} $-22$ & \cellcolor{newSCFT} & \cellcolor{newSCFT} \qquad $D_3^{24}(\mathfrak e_8)$ \\
\hhline{~---}

& $-\frac{29}{2}$ & & \\
\hline

\end{tabularx}

\caption{Simply-laced affine Kac--Moody algebras selected by determinant-based graded unitarity constraints. Entries in pale blue are either associated VOAs of rank-one Kodaira SCFTs, or are boundary admissible affine VOAs. Entries in pale green correspond to non-admissible affine VOAs conjectured to arise in the respective $D^b_q(\gf)$ AD theories.}
\label{tab:classification}

\end{table}

%% file: sections/sec_2_ADEalgebras.tex
%%%%%%%%
\section{Relevant aspects of the Gorelik--Kac determinant} \label{sec:GKvacdet}
%%%%%%%%
In this section, we review the Gorelik--Kac determinant of the universal vacuum module for simply-laced affine Kac--Moody algebras \cite{gorelik2007simplicity}. Following the simplified form presented in \cite{ArabiArdehali:2025fad}, the vacuum determinant for $\gf$ simply laced is given by, 
\begin{equation} \label{eq:KacGorelikvacdet}
    \text{det}~S_\nu(k) \propto (k+h^\vee)^{m_0(\nu)} \prod_{\substack{p \geqslant 2, q \geqslant 1 \\ (p,q) = 1}} (k-k_{p,q})^{m_{p/q}(\nu)}~,
\end{equation}
where $h^\vee$ is the dual Coxeter number of $\gf$ and the levels are defined by,
\begin{equation}
    k_{p,q} = - h^\vee + \frac{p}{q}~.
\end{equation}
The generating function of the exponent $m_0(\nu)$ is given by,
\begin{equation}
    \sum_{\nu \in \hat{\Delta}_+} m_0(\nu) e^{-\nu} = \frac{1}{\tilde{R}} \sum_{m = 2}^{h^\vee} \sum_{i = m}^\infty (s_{m-1} - s_m) \frac{e^{-i \delta}}{1 - e^{-i \delta}}~,
\end{equation}
where $\delta$ is the imaginary root of $\hat{\gf}$, $s_j := \# \{ \alpha \in \Delta_+ : (\rho|\alpha) = j \}$, $\rho$ is the Weyl vector of $\gf$, $\Delta_+$ is the set of positive roots of $\gf$, and
\begin{equation}
    \tilde{R} = \sum_{\gamma \in \hat{\Delta}_+ / \Delta_+} (1-e^{-\gamma})^{\text{dim} \hat{\gf}_\gamma}~.
\end{equation}
Here, $\hat{\Delta}_+ = \Delta_+ \sqcup \left\{\alpha + n\delta\,\middle|\,\alpha \in \Delta, \ n\geqslant 1 \right\} \sqcup \left\{n\delta \,\middle|\, n \geqslant 1 \right\}$ is the set of positive affine roots and dim$\hat{\gf}_\gamma$ = dim$\hf$ if $\gamma \in \mathbb{Z} \delta$, and dim$\hat{\gf}_\gamma$ = 1 if $\gamma \notin \mathbb{Z} \delta$. 

The generating function for the exponents $m_{p/q}(\nu)$ is given by,
\begin{equation} \label{eq:multAKMmain}
    \sum_{\nu \in \hat{\Delta}_+} m_{p/q}(\nu) e^{-\nu} = \frac{M_{p/q}}{R}~,
\end{equation}
where,
\begin{equation}
\begin{split}
    &M_{p/q} := \sum_{\alpha \in \Delta_+} \sum_{j \in \mathbb{Z}_{\geqslant 1}}^{\prime} e^{-(jp-(\rho|\alpha))jq \delta} E((jp - (\rho|\alpha))\alpha) - \sum_{\alpha \in \Delta_+} \sum_{j \in \mathbb{Z}_{\geqslant 1}} e^{-(jp+(\rho|\alpha))jq \delta} E(jp\alpha)~, \\ 
    &R := \prod_{\gamma \in \hat{\Delta}_+} (1-e^{-\gamma})^{\text{dim} \hat{\gf}_\gamma}~, \\ 
    &E(\lambda) := \sum_{w \in \text{Weyl}(\gf)} (-1)^{\ell(w)} e^{w(\lambda + \rho)-\rho}~.
\end{split}
\end{equation}
The prime on the first sum indicates that the sum is over all $j \in \mathbb{Z}_{\geqslant 1}$ satisfying $j p -(\rho|\alpha) \geqslant 1$. It was shown in \cite{gorelik2007simplicity} that the exponent $M_{p/q} = 0$ for $p = 1$ and $p < 0$. A key ingredient in computing the exponents $M_{p/q}$ explicitly is the following regularity criterion:
\begin{equation} \label{eq:ADEregularity}
    E(\lambda) \neq 0 \Longleftrightarrow \lambda+\rho \ \ \mathrm{is} \ \ \mathrm{regular}, \ \ i.e., \forall \ \alpha \in \Delta_+, \ (\lambda + \rho|\alpha) \neq 0~. 
\end{equation}
To keep track of the different contributions to $M_{p/q}$, it is convenient to decompose it as $M_{p/q} = M_{p/q}^+ - M_{p/q}^-$ where,
\begin{equation} \label{eq:splitM}
\begin{split}
    &M_{p/q}^+ = \sum_{\alpha \in \Delta_+} \sum_{\substack{t \in \mathbb{Z}_{\geqslant 1}: \\ p|(t+(\rho|\alpha))}} e^{-t(t+(\rho|\alpha))\frac{q} {p}\delta} E(t \alpha) =: \sum_{\nu \in \hat{\Delta}_+} a^+_\nu e^{-\nu}~, \\
    &M_{p/q}^- = \sum_{\alpha \in \Delta_+} \sum_{\substack{t \in \mathbb{Z}_{\geqslant 2}: \\ p|t}} e^{-t(t+(\rho|\alpha))\frac{q}{p}\delta} E(t \alpha)  =: \sum_{\nu \in \hat{\Delta}_+} a^-_\nu e^{-\nu}~.
\end{split}
\end{equation}
As with the notation used in \cite{ArabiArdehali:2025fad}, we denote the restriction of the exponent at a fixed conformal level $L$ by $M_{p/q,L} := M^{+}_{p/q,L} - M^{-}_{p/q,L}$ where $M^\pm_{p/q, L} = \sum_{\mu \in Q^+} a^\pm_{L\delta + \mu} e^{-\mu}$, with $Q^+$ denoting the positive root lattice of $\gf$.  

The computation of the multiplicities $M_{p/q,L}$ is facilitated by the determination of the various $(t,\alpha)$ appearing in \eqref{eq:splitM} with $E(t\alpha) \neq 0$. In particular, note that they must satisfy $B_{(t,\alpha)} := t(t+(\rho|\alpha)) \frac{q}{p} \in \mathbb{N}q$ and $B_{(t,\alpha)} \leqslant L$. 

The strategy used in determining the multiplicities of zeros in the vacuum determinant will be to first characterise the possible values of $A_{(t,\alpha)} := t(t+(\rho|\alpha))$ for which $E(t\alpha)\neq0$, note down the corresponding $(t,\alpha)$, and then use that to compute relevant multiplicities. 

The multiplicities of the zeros of the Gorelik--Kac vacuum determinant (in all charge sectors) for simply-laced affine Kac--Moody algebras are encoded in
\begin{equation}
    \frac{M_{p/q,L}}{E(0)} = \frac{M^+_{p/q,L}}{E(0)} - \frac{M^-_{p/q,L}}{E(0)}~,
\end{equation}
where $E(0) = \prod_{\alpha \in \Delta_+}(1-e^{-\alpha})$. In general, the regularity criterion \eqref{eq:ADEregularity} and the expression for the exponents in \eqref{eq:splitM} imply that these functions are expressible as sums of characters of representations of the respective Lie algebra $\gf$,\footnote{While each weight $\lambda = t\alpha$ appearing in the expressions \eqref{eq:splitM} is integral, it need \emph{not} be dominant. However, the regularity criterion in \eqref{eq:ADEregularity} implies that one can perform a series of Weyl reflections so that the weight obtained after the shifted Weyl action $w.\lambda = \Lambda$ is dominant, and as a consequence $E(\lambda) = (-1)^{\ell(w)} E(\Lambda)$.} 
\begin{equation} \label{eq:expinchar}
\begin{split}
    \frac{M_{p/q,L}}{E(0)} &= \sum_{\substack{\Lambda \ \mathrm{dominant} \ \mathrm{integral} }} s_\Lambda \frac{E(\Lambda)}{E(0)}\\ &=\sum_{\substack{\Lambda \ \mathrm{dominant} \ \mathrm{integral} }} s_\Lambda~\mathrm{ch}(\Lambda), \qquad s_\Lambda \in \mathbb{Z}~. 
\end{split}
\end{equation}
Here $\mathrm{ch}(\Lambda)$ is the character of the representation of $\gf$ with highest weight $\Lambda$. Therefore, determining the multiplicities of zeros in the Gorelik--Kac vacuum determinant in a fixed-weight sector at a fixed conformal level reduces to finding the $(\Lambda, s_\Lambda)$ appearing in the above function. As a simple example, note that for $k = k_{p,q}$ boundary admissible, \emph{i.e.}, $p = h^\vee$ and $(p,q) = 1$,
\begin{equation}
    \frac{M_{p/q,q}}{E(0)} = \frac{E(\theta)}{E(0)} = \mathrm{ch}~\mathrm{Ad}~,
\end{equation}
which implies that the conformal level $L = q$ primitive null state at $k = k_{p,q}$ is a component of the adjoint representation. 

%% file: sections/sec_3_graded_unitarity.tex
%%%%%%%%%%%
\section{Graded unitarity structures on affine Kac--Moody algebras} \label{sec:grunitstructures}
%%%%%%%%%%%
Any vertex algebra associated to a unitary four-dimensional $\mathcal{N}=2$ SCFT comes equipped with a graded unitary structure. This is a package of extra structures that consists of a (generally order-four) anti-linear conjugation $\rho$ as well as a (generally half-integral) nondegenerate \emph{good} filtration $\mfR_\bullet$. The $\mfR$-filtrations on the VOAs associated with interacting four-dimensional SCFTs must obey additional conditions, and were dubbed \textit{interacting $\mfR$-filtrations (of four-dimensional type)} in \cite{ArabiArdehali:2025fad}. 

We are interested in cases where the underlying vertex algebra is taken to be the affine Kac--Moody algebra $V_k(\gf)$, where $\gf$ is a semisimple Lie algebra. While the bulk of this paper is dedicated to the case where $\gf$ is simply laced, we make no such restriction here, and discuss the graded unitarity structures more generally. Since $V_k(\gf)$ is a $\mathbb{Z}$-graded bosonic vertex algebra, the graded-unitary version of the spin-statistics relation \cite{Beem:2025guj} requires that the $\mfR$-filtration be integral. As a consequence, the conjugation $\rho$ must be an involution.

Since $\rho$ is a $\mathbb{C}$-antilinear vertex algebra automorphism, its action on the vertex algebra is completely determined by its action on the strong generators $J^a$. Furthermore, the action of $\rho$ selects the compact real form, and with a suitable choice of basis, its action on the strong generators can be expressed as $\rho(J^a) = -J^a$. The determination of the full $\mathfrak{R}$-filtration on $V_k(\gf)$ from first principles is beyond the scope of this work. Instead, we \emph{assume} that the $\mathfrak{R}$-filtration is the one proposed in \cite{ArabiArdehali:2025fad},
\begin{equation} \label{eq:AKMRfiltration}
\scalebox{0.9}{$\displaystyle
    \mathfrak{R}_p V_k(\gf) = {\rm span}\Big\{ (\prod_{i=1}^{\text{rank}~\gf} C^{(s_i)}_{-l_1^{(i)}} \cdots C^{(s_i)}_{-l_{k_i}^{(i)}}) J^{a_1}_{-n_1} \cdots J^{a_k}_{-n_k}  \Omega~; l_j^{(i)} \geqslant s_i, n_j \geqslant 1; ~~k+\sum_{i=1}^{\text{rank}~\gf}(s_i-1) k_i \leqslant p \Big\}~,
    $}
\end{equation}
where $C^{(s_i)}$ are the Casimir operators of \cite{Bouwknegt:1992wg},\footnote{Throughout the paper, the normal ordering is taken to be the nested normal ordering.}
\begin{equation}
    C^{(s_i)}(z) = d_{a_1 a_2 \ldots a_{s_i}} (J^{a_1} J^{a_2} \cdots J^{a_{s_i}})(z), \qquad i = 1, 2, \ldots, \text{rank}~\gf~,
\end{equation}
where $s_i$ are the degrees of primitive Casimir invariants of $\gf$ and $d_{a_1a_2\ldots a_{s_i}}$ are primitive symmetric invariant tensors in $\gf$. Assuming that the $\mfR$-filtration is non-degenerate, one obtains the orthogonal decomposition of a fixed conformal weight subspace, 
\begin{equation} \label{eq:orthogdecomp}
    \left(V_k(\gf) \right)_h = \bigoplus_{R = 0}^h \left(V_k(\gf) \right)_{h,R}
\end{equation}
with the upper bound following from the Schur condition $h = R+j_1+j_2$, which implies $R \leqslant h$.

The $\mathfrak{R}$-filtration in \eqref{eq:AKMRfiltration} is then the one that assigns each Casimir operator $C^{(s_i)}$ to the filtration component $p = s_i - 1$. This $\mathfrak{R}$-filtration is the ``simplest" modification consistent with the Higgs branch relations coming from the 4d $\mathcal{N} = 2$ SCFT which have $V_k(\gf)$ as the associated VOAs. This standard $\mfR$-filtration is also the one that was implicitly used in the computations of Macdonald indices \cite{Gadde:2011uv} carried out in \cite{Song:2016yfd, kang2026indexnonpointrgflows, Kang:2026ney}.\footnote{See \cite{Jiang:2026ffv} for a discussion on using the $\mfR$-filtration and graded unitarity structures for computing specific limits of Macdonald indices.}  

%%%%%%%%%%
\subsection{Goodness of the standard $\mathfrak{R}$-filtration}
%%%%%%%%%%
The $\mathfrak{R}$-filtration on any vertex algebra associated to a 4d $\mathcal{N} = 2$ SCFT must satisfy the necessary property of \emph{goodness}, which follows from noting that the OPE of two Schur operators is non-singular. Additionally, the $\mfR$-filtration is increasing and exhaustive. Concretely, for affine Kac--Moody algebras, an integral, increasing, and exhaustive good filtration is \cite{LiVertexandVPA, ArabiArdehali:2025fad},
\begin{equation}
    \ldots \subseteq \mathcal{G}_{-1} \mathbb{V} \subseteq \mathcal{G}_{0} \mathbb{V} \subseteq \mathcal{G}_{1} \mathbb{V} \subseteq \ldots, \qquad \bigcup_{m \in \mathbb{Z}}^\infty \mathcal{G}_{m} \mathbb{V} = \mathbb{V}~,
\end{equation}
for which the identity $\Omega \in \mathcal{G}_0 \mathbb{V}$, and for any $a \in \mathcal{G}_{m}\mathbb{V}$ and $b \in \mathcal{G}_{n}\mathbb{V}$,
\begin{equation}
\begin{split}
    a_i b \in \mathcal{G}_{m+n} \mathbb{V}, \qquad i < 0~, \\
    a_i b \in \mathcal{G}_{m+n-1} \mathbb{V}, \ \ \  i \geqslant 0~.
\end{split}
\end{equation}
Taking the commutative limit (more precisely, the associated graded) of the VOA with respect to any good filtration yields a Poisson vertex algebra (PVA) \cite{LiVertexandVPA}. The PVA obtained by taking the commutative limit with respect to the $\mfR$-filtration corresponds to the PVA obtained by the holomorphic-topological twist of the parent 4d theory \cite{Kapustin:2006hi}. Conversely, the VOA can be viewed as a certain Omega deformation of the holomorphically twisted theory \cite{Oh:2019bgz, Jeong:2019pzg}. 

It was conjectured in \cite{ArabiArdehali:2025fad} that the standard $\mathfrak{R}$-filtration in \eqref{eq:AKMRfiltration} is \emph{good}. In the remainder of this section, we will show that this is indeed the case. It suffices to show that all operators appearing in the singular terms of the OPEs $J^a(z) C^{(s_i)}(0)$ and $C^{(s_i)}(z) C^{(s_j)}(0)$ are in the filtered subspaces  $\mathfrak{R}_{s_i-1} V_k(\gf)$ and $\mathfrak{R}_{s_i+s_j-3} V_k(\gf)$ respectively. To this end, note that the relevant OPEs have the following schematic form,
\begin{equation}
\begin{split}
    J^a(z) C^{(s_i)}(0) = \sum_{k = 1}^{s_i+1} \frac{\mathcal{A}_k}{z^k} + \text{regular terms}~, \\
    C^{(s_i)}(z) C^{(s_j)}(0) = \sum_{k = 1}^{s_i+s_j} \frac{\mathcal{B}_k}{z^k} + \text{regular terms}~.
\end{split}
\end{equation}
Here, $\mathcal{A}_k, \mathcal{B}_k$ are sums of derivatives of normal ordered products of the currents at the relevant pole order. Conformal covariance requires that $h(\mathcal{A}_k) = s_i-k+1 \leqslant s_i -1$ for all $k \geqslant 2$, and $h(\mathcal{B}_k) = s_i+s_j-k \leqslant s_i + s_j - 3$ for all $k \geqslant 3$. As a result, it follows from \eqref{eq:orthogdecomp} that $\mathcal{A}_{k \geqslant 2} \in \mathfrak{R}_{s_i - 1}V_k(\gf)$, and $\mathcal{B}_{k \geqslant 3} \in \mathfrak{R}_{s_i + s_j -3}V_k(\gf)$. Since the Casimir operators are constructed using invariant tensors, the order one pole term in the $J^aC^{(s_i)}$ OPE vanishes, implying $\mathcal{A}_{k \geqslant 1} \in \mathfrak{R}_{s_i - 1}V_k(\gf)$. Thus, proving goodness of the filtration reduces to showing that the operators appearing in the $C^{(s_i)} C^{(s_j)}$ OPE at the order one pole with \emph{zero} or \emph{one} derivatives, and order two pole with \emph{zero} derivatives belong to the filtered subspace $\mathfrak{R}_{s_i + s_j -3}V_k(\gf)$. 

It is convenient to consider another filtration that keeps track of derivatives: the Li filtration. All vertex algebras are equipped with the Li filtration, which is a decreasing, good filtration. For affine Kac--Moody algebras $V_k(\gf)$, this filtration is given by,
\begin{equation}
    \mathfrak{F}^p V_k(\gf)= \mathrm{span}\left\{ J^{a_1}_{-1-n_1} J^{a_2}_{-1-n_2} \ldots J^{a_m}_{-1-n_m} \Omega; \ \ \ n_i \geqslant 0; \ \ \ \sum_{i = 1}^{m} n_i \geqslant p \right\}~.
\end{equation}
The degree-zero subspace in the associated graded of $V_k(\gf)$ with respect to the Li filtration comprises sums of normal ordered products of currents without any derivatives, $\mathrm{gr}_0^{\mathfrak{F}^\bullet} V_k(\gf) \cong \mathfrak{F}^0V_k(\gf) / \mathfrak{F}^1V_k(\gf)$, and is also referred to as the Zhu $C_2$ algebra \cite{Zhu:1996gaq}. Similarly, the quotient algebra $\mathfrak{F}^1V_k(\gf) / \mathfrak{F}^2V_k(\gf)$ comprises sums of normal ordered products of operators with exactly one derivative. The strategy to proving the goodness of the $\mfR$-filtration in \eqref{eq:AKMRfiltration} is to use the Li filtration and reduce the problem to a question of invariant Lie theory. 

To begin with, consider the OPE of the currents to be given by,
\begin{equation}
    J^a(z) J^b(w) = \frac{k \delta^{a b}}{(z-w)^2} + \frac{i f^{ab}_{\ \ c} J^c(w)}{(z-w)} + \text{regular terms}~,
\end{equation}
where we have chosen the normalisation such that the non-degenerate Killing form on the Lie algebra $\gf$ is $\kappa^{ab} = \delta^{ab}$. We will also use the resulting identity,
\begin{equation} \label{eq:AntiSymJJ}
    (J^a J^b) - (J^bJ^a) = if^{ab}_{\ \ c} \partial J^c~. 
\end{equation}
Following standard techniques from \cite{Bais:1987dc, SEVRIN1988465, Kac:1996wd}, it can be shown by direct computation that the $\mathcal{O}(z^{-2})$ pole term is given by,
\begin{equation} \label{eq:order2pole}
\begin{split}
    C^{(s_i)}(z) C^{(s_j)}(0)|_{\mathcal{O}(z^{-2})} &= d_{a_1 \ldots a_{s_i}} d_{b_1 \ldots b_{s_j}}\left(k \sum_{m=1}^{s_i} \sum_{n=1}^{s_j} \delta^{a_m,b_n} \left(J^{\hat{A}_m} J^{\hat{B}_n}\right) \right. \\ 
    &\left. - \sum_{\substack{m=1}}^{s_i} \sum_{1 \leqslant n_1 < n_2}^{s_j} f^{a_{m} b_{n_1}}_{\qquad \ c} f^{c b_{n_2}}_{\ \ \ \ \ d} \left(J^d J^{\hat{A}_{m}} J^{\hat{B}_{(n_1,n_2)}}\right) \right) + \ldots~, 
\end{split}
\end{equation}
where the shorthand notations indicate,
\begin{equation}
\begin{split}
    &J^{\hat{A}_{m}} := J^{a_1} \cdots J^{a_{m-1}} J^{a_{m+1}} \cdots J^{a_{s_i}}, \qquad J^{\hat{B}_{n}} := J^{b_1} \cdots J^{b_{n-1}} J^{b_{n+1}} \cdots J^{b_{s_j}}~, \\
    &J^{\hat{B}_{(n_1,n_2)}} := J^{b_1} \cdots J^{b_{n_1-1}} J^{b_{n_1+1}} \cdots J^{b_{n_2-1}} J^{b_{n_2+1}} \cdots J^{b_{s_j}}~,
\end{split}
\end{equation}
and $\ldots$ represent sums of normal ordered products of operators with at least one derivative, and hence belong to $\mathfrak{F}^1 V_k(\gf)$. Since re-normal ordering of the operators introduces at least one more derivative, the tensor appearing in the coefficient of the normal ordered products can be completely symmetrised in all the indices without changing the $\mathcal{O}(z^{-2})$ pole term in the Zhu $C_2$ algebra. Therefore, the different terms can all be clubbed together and rewritten in the schematic form,
\begin{equation} \label{eq:order2simp}
    C^{(s_i)}(z) C^{(s_j)}(0)|_{\mathcal{O}(z^{-2})} = X^{(1)}_{(\alpha_1 \alpha_2 \ldots \alpha_{s_i+s_j-2})} (J^{\alpha_1} J^{\alpha_2} \ldots J^{\alpha_{s_i+s_j-2}}) + \ldots~,
\end{equation}
where $\ldots$ represent sums of normal ordered products of operators with at least one derivative, and hence belong to $\mathfrak{F}^1 V_k(\gf)$. The tensor $X^{(1)}_{(\alpha_1 \alpha_2 \ldots \alpha_{s_i+s_j-2})}$ is a fully symmetric invariant tensor, with the invariance following from the $G$-equivariance of the OPE, and is hence valued in the symmetric algebra $\left(S^{(s_i+s_j-2)} \gf^*\right)^{\gf}$. Therefore, the $\mathcal{O}(z^{-2})$ term in the OPE with no derivatives can be expressed as sums of normal ordered products of Casimir operators, leading to,
\begin{equation}
    C^{(s_i)}(z) C^{(s_j)}(0)|_{\mathcal{O}(z^{-2})} \in \mathfrak{R}_{s_i+s_j-3} V_k(\gf)~.
\end{equation}
Similarly, the $\mathcal{O}(z^{-1})$ pole takes the form,
\begin{equation} \label{eq:order1simp}
\begin{split}
    C^{(s_i)}(z) C^{(s_j)}(0)|_{\mathcal{O}(z^{-1})} &= \sum_{n=1}^{s_j} \sum_{m = 1}^{s_i} d_{b_1 \ldots b_{s_j}} d_{a_1 \ldots a_{m-1} c a_{m+1} \ldots a_{s_i}} if^{b_n c}_{\ \ \ \ a_m} (J^{B_{1}} J^{A} J^{B_{2}})(0) \\ &+ X^{(2)}_{a (\alpha_1 \alpha_2 \ldots \alpha_{s_i+s_j-3})} (\partial J^aJ^{\alpha_1} J^{\alpha_2} \cdots J^{\alpha_{s_i+s_j-3}})(0) + \ldots~.
\end{split}
\end{equation}
Here, we have again introduced a similar shorthand notation, 
\begin{equation}
\begin{split}
    &J^{B_{1}} = J^{b_1} J^{b_2} \cdots J^{b_{n-1}}~, \\
    &J^{A} = J^{a_1} J^{a_2} \cdots J^{a_{s_i}}~, \\
    &J^{B_{2}} = J^{b_{n+1}} J^{b_{n+2}} \cdots J^{b_{s_j}}~.
\end{split}
\end{equation}
The $\ldots$ represent operators that have at least two derivatives and are in $\mathfrak{F}_2 V_k(\gf)$. While the precise form of the second term in \eqref{eq:order1simp} can be exactly computed, it is not illuminating. The relevant key property is that $X^{(2)}_{a (\alpha_1 \alpha_2 \ldots \alpha_{s_i+s_j-3})}$ is an invariant tensor that is symmetric in the bracketed indices. As before, the symmetry follows from \eqref{eq:AntiSymJJ}, while the invariance follows from the $G$-equivariance of the OPE. 

The invariance of the tensor $d_{a_1 a_2 \ldots a_{s_i}}$ implies that $\sum_{m=1}^{s_i} f^{b c}_{\ \ a_{m}} d_{a_1 \ldots a_{m-1} c a_{m+1} \ldots a_{s_i}} = 0$, which leads to the vanishing of the first term in \eqref{eq:order1simp} above. Since it suffices to work at the level of the quotient algebra $\mathfrak{F}^1V_k(\gf)/\mathfrak{F}^2V_k(\gf)$, the problem can be recast as a question of invariant Lie theory. To this end, consider  
\begin{equation} \label{eq:F_adefn}
    F_a(j) := X_{a(\alpha_1 \alpha_2 \ldots \alpha_k)} j^{\alpha_1} j^{\alpha_2} \ldots j^{\alpha_k}~,
\end{equation}
where $X_{a(\alpha_1 \alpha_2 \ldots \alpha_k)}$ is an invariant tensor, symmetric in the bracketed indices, and $F_a(j)$ are elements in the symmetric algebra $S\gf^*$. Choosing a basis $\{e_a : a = 1,2,\ldots, \mathrm{dim}(\gf) \}$ for $\gf$ and the dual basis $\{e^a : a = 1,2,\ldots, \mathrm{dim}(\gf) \}$ for $\gf^*$, one can define $F(j) = \sum_{a=1}^{\mathrm{dim}(\gf)} F_a(j) e^a$, which is equivalently the linear map (defined on the basis of $\gf$ and extended linearly),
\begin{equation} 
\begin{split}
    F(j) &: \gf \longrightarrow S\gf^*~, \\
    &: e_a \mapsto F_a(j)~.  
\end{split}
\end{equation}
Furthermore, the invariance of the tensor $X_{a(\alpha_1 \alpha_2 \ldots \alpha_k)}$ implies that $F(j) \in \mathrm{Hom}_{\gf}(\gf, S\gf^*)$. 

By the Chevalley restriction theorem, the ring of $G$-invariant functions on $\gf$ is isomorphic to the Weyl-invariant polynomial ring on the Cartan $\mathfrak{h} \subset \gf$,
\begin{equation} \label{eq:Chevalley}
    \mathbb{C}[\gf]^G \cong \mathbb{C}[\mathfrak{h}]^W \cong \mathbb{C}[c_1,c_2,\ldots, c_{\mathrm{rank}(\gf)}]~,
\end{equation}
where $c_i$ are the independent casimir elements of $\gf$. According to Kostant's theorem \cite{Kostant:1963}, given a complex semisimple Lie algebra $\gf$, the symmetric Lie algebra $S\gf$ is a free module over the commutative ring $(S\gf)^\gf$. Moreover, for every finite dimensional representation $V$ of $\gf$, the space $\mathrm{Hom}_{\gf}(V, S \gf)$ is a free $(S\gf)^\gf$-module of rank $\mathrm{dim}(V[0])$, \emph{i.e.}, the dimension of the zero-weight space of $V$. 

The nondegenerate invariant bilinear form on $\gf$ leads to the identification $\gf \cong \gf^*$. Thus, Kostant's theorem with $V$ taken to be the adjoint representation of $\gf$, and \eqref{eq:Chevalley} together imply that $\mathrm{Hom}_{\gf}(\gf, S\gf^*)$ is a free module over $\mathbb{C}[c_1, c_2, \ldots, c_{\mathrm{rank}(\gf)}]$ of rank equal to the rank of the Lie algebra, rank$(\gf)$. Thus, $F(j)$ is expressible as,
\begin{equation} \label{eq:Kostant}
    F(j) = \sum_{i=1}^{\mathrm{rank}(\gf)} a_i f_i, \qquad \mathrm{with} \qquad a_i \in \mathbb{C}[c_1, c_2, \ldots, c_{\mathrm{rank}(\gf)}]~.
\end{equation}
In fact, a theorem attributed to Vust, identifies the basis $\{f_i\}$ with the differentials of the casimir elements, \emph{i.e.}, $f_i = dc_i$ \cite{Vust1974, PANYUSHEV2002125, panyushev2007semi}.\footnote{The author thanks Christopher Beem for helpful discussions on this topic.}

To explain this in more detail, note that the casimir elements $c_i \in (S\gf^*)^\gf$ can be equivalently regarded as polynomial maps $c_i : \gf \longrightarrow \mathbb{C}$, and the differentials of the casimir elements are then given by the linear maps $(dc_i)_x : T_{x} \gf \longrightarrow \mathbb{C}$ where $x \in \gf$. Since the tangent space $T_{x} \gf$ is isomorphic to $\gf$, the differentials are valued in $\gf^*$. Consequently, one obtains,
\begin{equation}
    (dc_i)_x = \sum_{a = 1}^{\mathrm{dim}(\gf)} \frac{\partial c_i}{\partial j^a} (x) dj^a = \sum_{a = 1}^{\mathrm{dim}(\gf)} \frac{\partial c_i}{\partial j^a} (x) e^a~. 
\end{equation}
When combined with \eqref{eq:Kostant}, this implies that $F(j)$ can be rewritten as,
\begin{equation}
    F(j) = \sum_{i=1}^{\mathrm{rank}(\gf)}\sum_{a = 1}^{\mathrm{dim}(\gf)} a_i \frac{\partial c_i}{\partial j^a} e^a,  \qquad \mathrm{with} \qquad a_i \in \mathbb{C}[c_1, c_2, \ldots, c_{\mathrm{rank}(\gf)}]~.
\end{equation}
Comparing this with the definition in \eqref{eq:F_adefn}, one finally obtains,
\begin{equation}
    X_{a(\alpha_1 \alpha_2 \ldots \alpha_k)} j^{\alpha_1} j^{\alpha_2} \ldots j^{\alpha_k} = \sum_{i=1}^{\mathrm{rank}(\gf)} \mathcal{F}_i(c_1,c_2,\ldots, c_{\mathrm{rank}(\gf)}) \frac{\partial c_i}{\partial j^a}~,
\end{equation}
where $\mathcal{F}_i$ is a polynomial function of the casimir elements. The restriction of the Casimir operators to the Zhu $C_2$ algebra is precisely given by the casimir elements as a polynomial function of the currents $J^a$. Consequently, one finds that,
\begin{equation}
    X^{(2)}_{a (\alpha_1 \ldots \alpha_{s_i+s_j-3})} (\partial J^aJ^{\alpha_1} \ldots J^{\alpha_{s_i+s_j-3}}) = \sum_{i=1}^{\mathrm{rank}(\gf)} \mathcal{F}_i\left(C^{(s_1)},\ldots, C^{(s_{\mathrm{rank}(\gf)})}\right) \partial C^{(s_i)} + \ldots~,
\end{equation}
where $\ldots$ are sums of normal ordered products of currents belonging to $\mathfrak{F}^2V_k(\gf)$, implying,
\begin{equation}
    C^{(s_i)}(z) C^{(s_j)}(0)|_{\mathcal{O}(z^{-1})} \in \mathfrak{R}_{s_i+s_j-3} V_k(\gf)~,
\end{equation}
thus establishing the goodness of the standard $\mathfrak{R}$-filtration for affine Kac--Moody algebras $V_k(\gf)$ with semisimple $\gf$. 

\subsection{Sign constraints from graded unitarity}
We now restrict our attention to affine Kac--Moody algebras $V_k(\gf)$ where $\gf$ is simply laced. Graded unitarity imposes constraints on the signs of most singular terms in the OPEs of $\mathcal{O}(z)$ and $\rho(\mathcal{O})(0)$ that are related to $\mathcal{O}$'s position in the $\mathfrak{R}$-filtration \cite{ArabiArdehali:2025fad}. In particular, writing the two-point function as
\begin{equation}
    \mathcal{O}(z) \rho(\mathcal{O})(0) = \frac{\kappa_{\mathcal{O}}}{z^{2 h_\mathcal{O}}}~,
\end{equation}
graded unitarity requires
\begin{equation}
    \kappa_\mathcal{O} \propto (-1)^{h_{\mathcal{O}} - R_{\mathcal{O}}}~,
\end{equation}
with the proportionality constant being positive. Here, we make the assumption that the $\mathfrak{R}$-filtration \eqref{eq:AKMRfiltration} is nondegenerate. Consequently, it can be refined to a unique $R$-grading by Gram-Schmidt, with the $R$-charge of the operator $\mathcal{O}(z)$ being $R_{\mathcal{O}}$. 

Determinant-based graded unitarity sign constraints for simply-laced affine Kac--Moody algebras were established in \cite{ArabiArdehali:2025fad}, and we only state the final result here. In particular, graded unitarity imposes the following sign constraint on the Gorelik--Kac vacuum determinant \eqref{eq:KacGorelikvacdet} at conformal level $L$ in a fixed weight $\nu$ sector,
\begin{equation} \label{eq:grunitsignrule}
    \mathrm{sign}\left(\mathrm{det}S_{L\delta+\nu} \right) = (-1)^{\sum_{\substack{p \geqslant 2, q \geqslant 1:} (p,q) = 1} m_{p/q}(L\delta + \nu)}~.
\end{equation}

For any interacting $\mfR$-filtration of four-dimensional type, $R(J) = R(T) = 1$. Since the stress tensor for $V_k(\gf)$ is given by the Segal--Sugawara stress tensor, graded unitarity requires $-h^\vee < k \leqslant 0$, and thus, the factors of $(k+h^\vee)^{m_0}$ in the vacuum determinant are positive. Hence, \eqref{eq:grunitsignrule} implies that when the vacuum determinant in a fixed-weight sector is non-vanishing, then only an \emph{even} number of non-critical factors $(k - k_{p,q})$ are positive.

Before implementing graded unitarity constraints for simply-laced affine Kac--Moody algebras, we briefly outline the general strategy. At a given conformal level $L$, we use the closed-form vacuum determinant \eqref{eq:KacGorelikvacdet} at a fixed weight $\nu$ to determine the \emph{structure of zeros}. More precisely, this involves determining the most negative and second most negative zeros (in certain cases, also the third most negative zero) of the vacuum determinant, along with their multiplicities. While we primarily consider the ``fixed-root sector" \emph{i.e.}, $\nu = \alpha$, where $\alpha$ is any root of the respective Lie algebra $\gf$, other weight sectors are also analogously examined at certain low levels. Determinant-based graded unitarity constraints from \eqref{eq:grunitsignrule} for a given conformal level $L$ and an appropriate fixed weight $\nu$ are then imposed to exclude intervals of levels $k$ incompatible with graded unitarity. In addition, a bespoke low-level analysis by passing to a representation basis instead of a weight basis facilitates the exclusion of certain isolated levels that survive the examined determinant-based constraints. Extending this procedure to all conformal levels $L$, we find that the closed-form vacuum determinant \eqref{eq:KacGorelikvacdet}, together with the graded unitarity sign constraint \eqref{eq:grunitsignrule}, provides a remarkably powerful criterion for excluding levels $k$ of all simply-laced affine Kac--Moody algebras, ultimately leading to the classification summarised in Table \ref{tab:classification}.

%% file: sections/sec_4_typeA_graded_unitarity.tex
%%%%%%%%%
\section{Graded unitarity constraints on $V_k(\mathfrak{sl}_N)$} \label{sec:typeAgrunit}
%%%%%%%%%
In this section, we investigate constraints imposed by graded unitarity on $V_k(\mathfrak{sl}_N)$ affine Kac--Moody algebras. These constraints are implemented at the level of the vacuum determinant in a fixed-weight sector, and hence are weaker than the full set of graded unitarity constraints on the vertex algebra. 

Such an analysis has already been carried out for $\gf = \slf_2, \slf_3$ and $\slf_4$ \cite{ArabiArdehali:2025fad}, where it was shown that graded unitarity constraints are very restrictive and rule out all levels except,
\begin{equation}
\begin{split}
    &k = k_{h^\vee,q} = -h^\vee + \frac{h^\vee}{q}: (h^\vee,q) = 1, \qquad \mathrm{for} \ \ \gf = \mathfrak{sl}_2, \mathfrak{sl}_3~, \\
    &k = k_{4,q} = -4 + \frac{4}{q} \ \ \ \ \ :4 \nmid q, \qquad \qquad \ \ \ \mathrm{for} \ \ \gf = \mathfrak{sl}_4~.
\end{split}
\end{equation}
Note that for $\gf = \mathfrak{sl}_4$, there is no coprimality condition and it was shown \emph{loc. cit.} that, in addition to the boundary admissible levels, fixed root determinant-based constraints do not rule out the non-admissible $k = k_{2,q}$ levels with $q$ odd. We show that this pattern generalises to $\gf = \mathfrak{sl}_N$ for all $N$. 

In what follows, we restrict our attention exclusively to $\gf = \slf_N$ with $N \geqslant 5$. The positive roots and the Weyl vector for $\slf_N$ are given by,
\begin{equation}
\begin{split}
    &\Delta_+ = \{\alpha_{r,s} := e_r - e_s: 1 \leqslant r < s \leqslant N \}, \\
    &\rho = \frac{1}{2} \sum_{\alpha \in \Delta_+} \alpha = \left(\frac{N-1}{2}~, \frac{N-3}{2}, \ldots, -\frac{(N-1)}{2} \right)~.
\end{split}
\end{equation}
For our purposes, it is relevant to simplify the regularity criterion \eqref{eq:ADEregularity} into something more amenable to calculations in $\gf = \slf_N$. The weights $\lambda$ appearing in the argument of $E(\lambda)$ in the exponent $M_{p/q}$ are always positive integral multiples of positive roots. Consequently, it can be shown by direct calculation that for $\gf = \slf_N$ and $\lambda = t \alpha_{r,s}$ (where $t \in \mathbb{Z}_{\geqslant 1}$ and $\alpha_{r,s} \in \Delta_+$), the regularity criterion simplifies to,\footnote{This result is in agreement with the simplification obtained in \cite{jiang2025singular}.}
\begin{equation} \label{eq:typeAregularity}
    E(t \alpha_{r,s}) \neq 0 \Longleftrightarrow t \geqslant r \ \ \mathrm{and} \ \ t \geqslant N - s+1~.
\end{equation}
The approach to implementing graded unitarity constraints is analogous to \cite{ArabiArdehali:2025fad}. For a given conformal level $L$, we first determine the smallest values of $A_{(t,\alpha)} := t(t+(\rho|\alpha))$ such that $E(t\alpha) \neq 0$ in \eqref{eq:splitM}. Since $\frac{A_{(t,\alpha)}}{L} \leqslant \frac{p}{q}$, the smallest values of $A_{(t,\alpha)}$ determine the most negative and the second most negative zero of the vacuum determinant in a fixed weight $\nu = \alpha$ sector, where $\alpha$ is a root of $\slf_N$. The multiplicities of these zeros are then found using \eqref{eq:expinchar}.

To this end, note that the simplified $\slf_N$ regularity criterion \eqref{eq:typeAregularity} implies that $(\rho|\alpha) \geqslant N - 2t + 1$. Consequently, one has $A_{(t,\alpha)} \geqslant t(N-t+1)$, leading to,
\begin{equation}
\begin{split}
    A_{(t,\alpha)} \begin{cases}
        = N \qquad \qquad \ \ \ \mathrm{when} \ \ t = 1 \ \ \mathrm{and} \ \ \alpha = \theta \\
        \geqslant (2N-2) \qquad \mathrm{otherwise}
    \end{cases}
\end{split}
\end{equation}
Thus, when $N \nmid L$, the $k = k_{N,L}$ zero in the fixed-root vacuum determinant at conformal level $L$ is a consequence of the primitive null state in the adjoint representation. The multiplicity of this zero in the vacuum determinant $\mathrm{det}~S_{L \delta + \alpha}$ can be computed from,
\begin{equation}
    \frac{M_{N/L,L}}{E(0)} = \frac{E(\theta)}{E(0)} = \mathrm{ch~Ad}~,
\end{equation}
to be equal to one. It is worth stressing that this result does \emph{not} depend on any coprimality condition between $N$ and $L$. To summarise, the non-critical zeros of the fixed root, conformal level $L$ vacuum determinant det $S_{L \delta + \alpha}$ are given by,
\begin{equation} \label{eq: typeAprelimzero}
    k = \begin{cases}
        k_{N,L} = -N + \frac{N}{L}: N \nmid L \\ 
        k_{N,q} = -N + \frac{N}{q}: q < L \ \ \mathrm{and} \ \ N \nmid q \\
        k_{A,B} = -N + \frac{A}{B}: A \geqslant 2N-2 \ \ \mathrm{and} \ \ B \leqslant L
    \end{cases}
\end{equation}

\subsection{Structure of zeros of the vacuum determinant}
Before discussing the precise structure of zeros of the fixed-root vacuum determinant, it is worthwhile to perform a bespoke analysis at $L = 1$ and $L = 2$. Note that the fixed-root vacuum determinant for $\gf = \slf_N$ at these levels is given by,
\begin{equation} \label{eq:lowleveltypeAdet}
\begin{split}
    &\mathrm{det}~S_{\delta + \alpha} \propto (k-k_{N,1}) \\
    &\qquad \qquad \propto \textcolor{blue}{(k - 0)}~, \\
    &\mathrm{det}~S_{2\delta + \alpha} \propto (k - k_{N,2})(k-k_{N-1,1})^{N-3} (k - k_{N,1})^{m_{N/1}(2\delta + \alpha)} (k - k_{N+1,1})^{m_{(N+1)/1}(2\delta + \alpha)}\\ 
    &\qquad \qquad  \propto \textcolor{blue}{\left(k + \frac{N}{2} \right)(k + 1)^{N-3}} (k - 0)^{m_{N/1}(2\delta + \alpha)} (k - 1)^{m_{(N+1)/1}(2\delta + \alpha)}~.
\end{split}
\end{equation}
Hence, from graded unitarity constraints at level one, and from the fact that the Segal--Sugawara stress tensor has $R$-charge $R(T) = 1$, one obtains that $-N < k \leqslant 0$. Furthermore, graded unitarity constraints at level two additionally rule out $k \in \left(-\frac{N}{2},-1\right)$. The $k = -N/2$ zero is a consequence of a primitive null state in the adjoint representation, while the zero at $k = -1$ comes from a state in the $[0,1,0,\ldots,1,0]$ representation. Passing to the adjoint representation basis instead of a fixed weight basis, the norm sign incompatibility with graded unitarity, of the state in the adjoint representation persists in the whole regime $k \in \left(-\frac{N}{2}, 0\right)$. Consequently, at level two, constraints from graded unitarity restrict the level $k$ to be in the regime $-N < k \leqslant -\frac{N}{2}$ or $k = 0$, where the latter corresponds to the trivial vertex algebra.

To implement graded unitarity constraints for $L \geqslant 3$, we will use \eqref{eq: typeAprelimzero} to find the most negative and the second most negative zeros in the vacuum determinant in the fixed-root sector. Consider a conformal level $L$ such that $N \nmid L$. From \eqref{eq: typeAprelimzero}, the most negative zero of the vacuum determinant is $k = k_{N,L}$. To find the second most negative zero, we divide the analysis into two cases,
\begin{itemize}
    \item Case 1: $N \nmid (L-1)$ \\ 
    \\
    Since $N \nmid (L-1)$, $k = k_{N,L-1}$ first appears as a zero of the vacuum determinant at conformal level $L-1$ with multiplicity one, and hence, is also a zero of the vacuum determinant at level $L$. For $N \geqslant 5$ and  $L \geqslant 3$, one obtains,
    \begin{equation} \label{eq:typeA2ndsmallest}
    \begin{split}
        &(2N-2)(L-1) - NL = (N-2)L - (2N-2) > 0 \\
        &\Longrightarrow k_{2N-2,L} = -N + \frac{2N-2}{L} > -N + \frac{N}{L-1} = k_{N,L-1}~.
    \end{split}
    \end{equation}
    This shows that the zeros of the third kind in \eqref{eq: typeAprelimzero} are always greater than $k = k_{N,L-1}$ and hence for conformal levels $L$ such that $N \nmid (L-1)$, the second most negative zero is $k = k_{N,L-1}$. 
    \item Case 2: $N | (L-1)$ \\ 
    \\
    In this case, $N \nmid (L-2)$. Thus, $k = k_{N,L-2}$ first appears as a zero of the vacuum determinant at conformal level $L-2$ with multiplicity one, and hence, is also a zero of the vacuum determinant at level $L$. For $N \geqslant 5$ and $N | (L-1)$, one obtains,
    \begin{equation}
    \begin{split}
        &(2N-2)(L-2) - NL = (N-2)L - (4N-4) > 0 \\
        &\Longrightarrow k_{2N-2,L} = -N + \frac{2N-2}{L} > -N + \frac{N}{L-2} = k_{N,L-2}~.
    \end{split}
    \end{equation}
    This shows that the second most negative zero of the vacuum determinant is $k = k_{N, L-2}$. 
\end{itemize}
Finally, for $L$ that is a multiple of $N$, \emph{i.e.}, $N | L$, the most negative zero is given by $k = k_{N, L-1}$ by an argument analogous to the one leading to \eqref{eq:typeA2ndsmallest}. 

To summarise, we obtain the following structure of zeros for $N \geqslant 5$ and $L \geqslant 3$, 
\begin{itemize}
    \item At level $L \neq (0,1)~\mathrm{mod}~N$, the most negative zero of the fixed-root vacuum determinant is given by $k = k_{N,L}$ and this zero always occurs with multiplicity one. Moreover, the second most negative zero is given by $k = k_{N,L-1}$. 
    \item At level $L = 0~\mathrm{mod}~N$, the most negative zero of the fixed-root vacuum determinant is given by $k = k_{N,L-1}$. 
    \item At level $L = 1~\mathrm{mod}~N$, the most negative zero of the fixed-root vacuum determinant is given by $k = k_{N,L}$ and this zero always occurs with multiplicity one. Moreover, the second most negative zero is given by $k = k_{N,L-2}$. 
\end{itemize}

\subsection{Constraints on the level}
The conclusion of the structure of zeros of the previous subsection is that the fixed-root vacuum determinant at level $L \geqslant 3$ takes the following form, \\
\\
\underline{\textbf{Case 1: } $L \neq 0~\mathrm{mod}~N$}
\begin{equation} \label{eq:typeAvacdet1}
\begin{split}
    &\mathrm{det}~S_{L\delta+\alpha} \propto \textcolor{blue}{(k-k_{N,L})^1}(k-k_{N,L-1})^{\#_1}( \ldots \ldots), \qquad L \neq 1~\mathrm{mod}~N~, \\
    &\mathrm{det}~S_{L\delta+\alpha} \propto \textcolor{blue}{(k-k_{N,L})^1}(k-k_{N,L-2})^{\#_2}( \ldots \ldots), \qquad L = 1~\mathrm{mod}~N~,
\end{split}
\end{equation}
where $( \ldots \ldots)$ denote other zeros of the vacuum determinant that are strictly greater (less negative) than $k = k_{N,L-1}$ (or $k_{N,L-2}$). $\#_1$ and $\#_2$ are some positive integers, and the proportionality factor is a $k$-independent rational number. 
\\
\\
\underline{\textbf{Case 2: } $L = 0~\mathrm{mod}~N$}
\begin{equation} \label{eq:typeAvacdet2}
    \mathrm{det}~S_{L\delta+\alpha} \propto \textcolor{blue}{(k-k_{N,L-1})^{\#_3}}( \ldots \ldots)~,
\end{equation}
where again $( \ldots \ldots)$ indicate zeros strictly greater than $k = k_{N,L-1}$ and $\#_3$ is some positive integer.\footnote{Throughout the paper, we adopt the convention that any omitted zeros in the relevant vacuum determinant are strictly greater than those displayed, $\#$ denotes a positive integer, and the proportionality factors are positive.} 

The all-level determinant-based constraints from graded unitarity now follow almost immediately. In particular, let us first consider $L \neq 0~\mathrm{mod}~N$. From \eqref{eq:typeAvacdet1}, we see that the vacuum determinant flips sign at $k = k_{N,L}$, with the next zero being at $k_{N,L-1}$ or $k_{N,L-2}$. Hence, at level $L \neq 0~\mathrm{mod}~N$, the graded unitarity sign rule \eqref{eq:grunitsignrule} eliminates the regime $k \in (k_{N,L}, k_{N,L-1})$ or $k \in (k_{N,L}, k_{N,L-2})$ (depending on whether $L \neq 1~\mathrm{mod}~N$ or $L = 1~\mathrm{mod}~N$), respectively. 

Turning to the $L = 0~\mathrm{mod}~N$ case, note that the most negative zero is $k = k_{N,L-1}$, and all the $k_{N,q}$ zeros (with $1 < q \leqslant L-1$ and $N \nmid q$) of the level $L-1$ vacuum determinant are also zeros of the level $L$ vacuum determinant. Since $L-1 \neq 1~\mathrm{mod}~N$, graded unitarity constraints already rule out the region $(k_{N,L-1}, k_{N,L-2})$ at conformal level $L-1$. Thus, we do not get any new constraints from graded unitarity at level $L = 0~\mathrm{mod}~N$. 

Therefore, we find that with the assumption of the standard $\mfR$-filtration on the $V_k(\mathfrak{sl}_N)$ affine Kac--Moody algebras, the all-level (fixed root) determinant-based graded unitarity constraints rule out all values of the level $k$ other than,
\begin{equation} \label{eq:typeAalllevel}
    k = \begin{cases}
        \textcolor{knownSCFTtext}{k_{N,q} = -N + \frac{N}{q}: (N,q) = 1} \\
        k_{N,q} = -N + \frac{N}{q}: (N,q) > 1 \ \ \mathrm{and} \ \ N \nmid q
    \end{cases}
\end{equation}
These constraints are a natural generalisation of the $\gf = \mathfrak{sl}_2, \mathfrak{sl}_3$, and $\mathfrak{sl}_4$ exclusions of \cite{ArabiArdehali:2025fad}, and demonstrate that the determinant-based constraints are independent of whether $N$ and $q$ are coprime. The first set of levels above is the set of boundary admissible levels. The $V_k(\slf_N)$ affine Kac--Moody algebras corresponding to these levels are precisely the ones associated with the ``class I" $D_q^N(\mathfrak{sl}_N)$ Argyres--Douglas SCFTs under the SCFT/VOA correspondence \cite{Xie:2016evu, Song:2017oew}. 

On the other hand, the second set of levels are all non-admissible. These appear as the levels of the $\mathfrak{sl}_N$ affine Kac--Moody subalgebras of the associated VOAs of $D_q^N(\mathfrak{sl}_N)$ Argyres--Douglas SCFTs having extra mass parameters/flavour symmetries coming from the irregular puncture; hence the VOAs associated with these theories are strictly \emph{larger} than $V_k(\mathfrak{sl}_N)$ \cite{Xie:2016evu, Song:2017oew, Beem:2023ofp}. While allowed by all-level (fixed root) determinant-based graded unitarity constraints (under the assumption of the standard $\mathfrak{R}$-filtration), we leave open the possibility that an analogous analysis in a different weight sector or a finer state-based graded unitarity analysis might rule them out. 

As an implication of \eqref{eq:typeAalllevel}, note that when $N$ is prime, determinant-based graded unitarity constraints are sufficient to rule out \textit{all} levels other than the boundary admissible $k = k_{N,q}$, precisely the ones associated with the ``class I" $D_q^N(\mathfrak{sl}_N)$ Argyres--Douglas SCFTs. 

%% file: sections/sec_5_typeD_graded_unitarity.tex
%%%%%%%%%
\section{Graded unitarity constraints on $V_k(\mathfrak{so}_{2N})$} \label{sec:typeDgrunit}
%%%%%%%%%
In this section, we investigate constraints imposed by graded unitarity on $V_k(\mathfrak{so}_{2N})$ affine Kac--Moody algebras with $N \geqslant 4$. The strategy is similar to the previous section, however, as we elaborate in this section, the analysis is technically complicated due to the intricate structure of zeros of the Gorelik--Kac vacuum determinant. 

We begin by recalling that the positive roots and the Weyl vector for $\sof_{2N}$ are given by,
\begin{equation}
\begin{split}
    &\Delta_+ = \{\alpha^\pm_{r,s} := e_r \pm e_s: 1 \leqslant r < s \leqslant N \}, \\
    &\rho = \frac{1}{2} \sum_{\alpha \in \Delta_+} \alpha = \left(N-1, N-2, \ldots, 1,0 \right),
\end{split}
\end{equation}
with $(\rho|\alpha_{r,s}^+) = 2N - r - s$ and $(\rho|\alpha_{r,s}^-) = s - r$.

As in the type $A$ affine Kac--Moody case, we first simplify the regularity criterion \eqref{eq:ADEregularity} for $\lambda = t \alpha$ where $t \in \mathbb{Z}_{>0}$ and $\alpha \in \Delta_+$. It can be shown by direct computation that for $\gf = \sof_{2N}$ and $\lambda = t \alpha^{\pm}_{r,s}$, the regularity criterion simplifies to,
\begin{equation} \label{eq:typeDregularity}
\begin{split}
     &E(t \alpha^+_{r,s}) \neq 0 \Longleftrightarrow \begin{cases}
         t \geqslant s \qquad \text{or} \\ 
         t \geqslant r \ \ \text{and} \ \ t = s - r
     \end{cases} \\
     &E(t \alpha^-_{r,s}) \neq 0 \Longleftrightarrow \begin{cases}
         t \geqslant r \ \ \text{and} \ \ t \geqslant 2N - s \qquad \text{or} \\ 
         t \geqslant r \ \ \text{and} \ \ t = 2N - 2s \qquad \text{or} \\ 
         t \geqslant r \ \ \text{and} \ \ t = 2N-r-s
     \end{cases}     
\end{split}
\end{equation}
Following a similar approach as in the previous section, we first determine the smallest values of $A_{(t,\alpha)} = t(t+(\rho|\alpha))$ consistent with the simplified regularity constraints in \eqref{eq:typeDregularity} to be,
\begin{equation} \label{eq:AtconsttypeD}
\begin{split}
    &A_{(t,\alpha^+)} \ \ \begin{cases}
        = (2N-2) \qquad \text{when} \ \ t = 1 \ \ \text{and} \ \ \alpha^+ = \alpha^+_{1,2} = \theta \\ \geqslant (4N-8) \qquad \text{otherwise}
    \end{cases} \\
    &A_{(t,\alpha^-)} \ \ \begin{cases}
        = (2N-2) \qquad \text{when} \ \ t = 2 \ \ \text{and} \ \ \alpha^- = \alpha^-_{2,N-1} \\
        = 2N \qquad \qquad \ \text{when} \ \ t = 2 \ \ \text{and} \ \ \alpha^- = \alpha^-_{1,N-1} \\ \geqslant (4N-8) \qquad \text{otherwise}
    \end{cases}
\end{split}
\end{equation}

As a direct consequence of \eqref{eq:AtconsttypeD}, it can be observed that $E(\theta)$ and $E(2\alpha^-_{2,N-1})$ are relevant for computing the multiplicity of the zero $k = k_{2N-2,L}$, while $E(2\alpha^-_{1,N-1})$ is relevant for computing multiplicity of the $k = k_{2N,L}$ zero in the conformal level $L$ vacuum determinant in a fixed-weight sector. The multiplicities of the zeros $k = k_{2N-2,L}$ and $k = k_{2N,L}$ can then be determined by tracking the possible occurrence of corresponding pairs $(t,\alpha)$ in the summation in \eqref{eq:splitM}. Thus using \eqref{eq:expinchar}, for $N \geqslant 4$, we find that the multiplicity of the $k = k_{2N-2,L}$ zero in the fixed root, conformal level $L$ vacuum determinant is given by,\footnote{We note that while the weights $2\alpha^-_{2,N-1}$ and $2\alpha^-_{1,N-1}$ are not dominant, it can be shown that $(s_N s_{N-1}s_1).(2\alpha^-_{2,N-1}) = \theta$ and $(s_N s_{N-1}).(2\alpha^-_{1,N-1}) = 2\omega_1$. Thus, $E(2\alpha^-_{2,N-1}) = -E(\theta)$ and $E(2\alpha^-_{1,N-1}) = E(2\omega_1)$.} 
\begin{equation} \label{eq:Lmultofk2Nm2}
\begin{split}
    &N \ \text{even}: \qquad m_{(2N-2)/L}(L\delta+\alpha) = \begin{cases}
        1 \qquad L \in 2\mathbb{N}+1 \ \ \text{and} \ \ L \neq 0~\text{mod}~(N-1) \\
        2 \qquad L \in 2\mathbb{N}+1 \ \ \text{and} \ \ L = 0~\text{mod}~(N-1) \\
        0 \qquad L \in 2\mathbb{N}
    \end{cases} \\
    &N \ \text{odd}: \qquad m_{(2N-2)/L}(L\delta+\alpha) = \begin{cases}
        1 \qquad L \in 2\mathbb{N}+1  \\
        1 \qquad L = (N-1)~\text{mod}~(2N-2) \\
        0 \qquad \text{otherwise}
    \end{cases}
\end{split}
\end{equation}
In the same manner, for $N \geqslant 5$, the multiplicity of the $k = k_{2N,L}$ zero in the fixed-root vacuum determinant at conformal level $L$ is given by,
\begin{equation} \label{eq:Lmultofk2N}
   m_{2N/L}(L\delta+\alpha) = \begin{cases}
        1 \qquad L \in 2\mathbb{N} \ \ \text{and} \ \ L \neq 0~\text{mod}~N \\
        0 \qquad \text{otherwise}
    \end{cases}
\end{equation}
To understand the structure of zeros of the all-level vacuum determinant, we note from \eqref{eq:AtconsttypeD}, \eqref{eq:Lmultofk2Nm2}, and \eqref{eq:Lmultofk2N}, that the (possible) zeros of the conformal level $L$ vacuum determinant in a fixed-root sector for $V_k(\sof_{2N})$ can be classified as,
\begin{equation} \label{eq: so2nprelimzero}
    k = \begin{cases}
        k_{2N-2,q} = -(2N-2) + \frac{2N-2}{q}: q \leqslant L, \ q \neq 0~\mathrm{mod}~(N-1), \ \ \text{and} \ \ q \ \mathrm{odd} \\ 
        k_{2N-2,q} = -(2N-2) + \frac{2N-2}{q}: q \leqslant L, \ q = (N-1)~\mathrm{mod}~(2N-2) \\
        k_{2N,q} = -(2N-2) + \frac{2N}{q} \qquad: q \leqslant L, \ q \neq 0~\mathrm{mod}~N, \ \ \text{and} \ \ q \ \mathrm{even} \\
        k_{A,B} = -(2N-2) + \frac{A}{B} \qquad \ \ : A \geqslant (4N-8) \ \ \text{and} \ \ B \leqslant L
    \end{cases}
\end{equation}
As a consequence of the multiplicities in \eqref{eq:Lmultofk2Nm2}, the structure of zeros and the constraints from graded unitarity are qualitatively different for even and odd $N$. To illustrate the general strategy, we begin with a detailed analysis of $V_k(\mathfrak{so}_8)$, before treating the general case of even $N$. The corresponding analysis for odd $N$ is deferred to Appendix \ref{app:VkSOodd}, and we conclude the section by summarising graded unitarity constraints for all $N \geqslant 5$. 

\subsection{$V_k(\sof_8)$}
Implementing graded unitarity constraints on $V_k(\sof_8)$ requires a meticulous study of multiplicities of zeros in the vacuum determinant in a fixed-weight sector, and the representations corresponding to the specific states becoming null at these zeros. We will first outline the structure of zeros of the fixed-root vacuum determinant at conformal levels $L \leqslant 6$, and then infer the constraints obtained by graded unitarity. This bespoke analysis will orient us for investigating all-level determinant-based graded unitarity constraints by invoking the sign constraint in \eqref{eq:grunitsignrule}.

\subsubsection{Low level graded unitarity constraints}
The vacuum determinant in a fixed-root sector for conformal levels $L = 1$ and $L = 2$ takes the following form, 
\begin{equation} \label{eq:lowlevelso8det1to2}
\begin{split}
    &\text{det}~S_{\delta + \alpha} \propto (k-k_{6,1}) \\
    &\qquad \qquad \propto \textcolor{blue}{(k - 0)}~, \\
    &\text{det}~S_{2\delta + \alpha} \propto (k - k_{4,1})^3(k-k_{6,1})^{\#_1} (\ldots \ldots)\\ 
    &\qquad \qquad  \propto \textcolor{blue}{(k + 2)^3}(k - 0)^{\#_1} (\ldots \ldots)~. \\ 
\end{split}
\end{equation}
Level one constraints combined with the requirement that $R(T) = 1$ immediately implies that $-6 < k \leqslant 0$. 
At $L = 2$, the multiplicity three of $(k+2)$ in a fixed-root sector is a manifestation of the triality symmetry of $\sof_8$. In particular, it can be shown that the zero at $k = -2$ is a consequence of three null states, which are components of the $[2,0,0,0]$, $[0,0,2,0]$ and $[0,0,0,2]$ representations. Consequently, one sees that the graded unitarity sign rule eliminates $k \in (-2,0)$, leading to the constraint $-6<k \leqslant -2$ or $k = 0$. 

The constraints from graded unitarity are more interesting and involved at $L \geqslant 3$. To this end, note that the vacuum determinant in a fixed-root sector at $L = 3,4,5,6$ is given by,
\begin{equation} \label{eq:lowlevelso8det3to6}
\begin{split}
    &\text{det}~S_{3\delta + \alpha} \propto (k - k_{2,1})^2 (k-k_{4,1})^{\#_2} (\ldots \ldots)\\ 
    &\qquad \qquad  \propto \textcolor{blue}{(k + 4)^2} (k +2 )^{\#_2} (\ldots \ldots)~,\\
    &\text{det}~S_{4\delta + \alpha} \propto (k - k_{2,1})^{32}(k-k_{7,2})^5(k-k_{4,1})^{\#_3} (\ldots \ldots)\\ 
    &\qquad \qquad  \propto \textcolor{blue}{(k + 4)^{32}(k + 5/2)^5}(k+2)^{\#_3} (\ldots \ldots)~, \\
    &\text{det}~S_{5\delta + \alpha} \propto (k - k_{6,5})(k-k_{2,1})^{322} (k-k_{7,2})^{\#_4} (\ldots \ldots)\\ 
    &\qquad \qquad  \propto \textcolor{blue}{(k+24/5)(k + 4)^{322}}(k + 5/2)^{\#_4} (\ldots \ldots)~, \\
    &\text{det}~S_{6\delta + \alpha} \propto (k - k_{6,5})^{\#_5}(k-k_{4,3})^3 (\ldots \ldots)\\ 
    &\qquad \qquad  \propto \textcolor{blue}{(k+24/5)^{\#_5}(k + 14/3)^3} (\ldots \ldots)~. \\
\end{split}
\end{equation}
First, note that at $L = 3$, there are two primitive null states at $k = -4$ and both are components of the adjoint representation, hence contributing to the multiplicity two in a fixed-root sector. As a result, \textit{even though} there is a \textit{new smallest zero} at $k = -4$ in the vacuum determinant at $L = 3$, there are no new fixed root determinant-based constraints from the graded unitarity sign rule \eqref{eq:grunitsignrule}. This is a first of many instances for $V_k(\sof_{2N})$ where the structure of zeros is qualitatively different from that encountered in the $V_k(\mathfrak{sl}_N)$ graded unitarity analysis. 

At $L = 4$, there is a single primitive null state at $k = -5/2$, which is a component of the $[0,2,0,0]$ representation. The multiplicity of five then follows from the root multiplicity of the $V_{2\omega_2}$ representation being five. Hence, at $L = 4$, graded unitarity constraints rule out $k \in (-5/2,-2)$. At $L = 5$, the primitive null state giving rise to the $k = -24/5$ zero is a component of the adjoint representation, and graded unitarity further rules out $k \in (-24/5,-4) \bigcup (-4, -5/2)$. At $L = 6$ there are no new constraints, and hence determinant-based graded unitarity constraints up to $L = 6$ restrict the level $k$ to be,
\begin{equation}
    k \leqslant -\frac{24}{5} \qquad \text{or} \qquad k = \left\{-4, -\frac{5}{2},-2,0 \right\}~.
\end{equation}
Extending these multiplicity computations to higher conformal levels, we find that fixed root determinant-based graded unitarity constraints up to $L = 12$ are given by,
\begin{equation} \label{eq:lowlevelso8fullconst}
    k \leqslant -\frac{60}{11} \qquad \text{or} \qquad k = \left\{-\frac{16}{3}, -\frac{26}{5}, -\frac{36}{7}, -\frac{24}{5}, -4, -\frac{5}{2}, -2, 0  \right\}~.
\end{equation}
The key feature of this analysis is that the constraints behave differently for $L=(1,2)~\mathrm{mod}~6$ and $L=(3,4,5)~\mathrm{mod}~6$, owing to the contrasting multiplicities of the relevant zeros. This distinction will play an important role in the simplification of the all-level analysis below, and is a recurring feature for $V_k(\sof_{2N})$ with $N$ even.

\subsubsection{Structure of zeros of the vacuum determinant}
It is useful to assemble the smallest conformal level $L_s$ at which $k = k_{2,q}, k_{3,q}$ and $k_{6,q}$ first appear as zeros of the fixed-root vacuum determinant, and the multiplicities of these zeros (which can be computed using \eqref{eq:expinchar}). We document the relevant data in Table \ref{tb:multandLevforso8}.

%%%%%%%%%%%%%%%%%%%%%%%%%%%
\begin{table}[]
\centering
\begin{tabular}{|c|c|c|}
\hline\hline
$k$ & $L_s$ & $m_{p/q}(L_s\delta+\alpha)$ \\
\hline\hline
$k_{2,q}$ & $3q$ & $2$ \\
$k_{3,q}$ & $7q$ & $40$ \\
$k_{6,q}$ & $q$ & $1$ \\
\hline
\end{tabular}
\caption{Smallest conformal level $L_s$, and multiplicities for different values of $k$.}
\label{tb:multandLevforso8}
\end{table}
%%%%%%%%%%%%%%%%%%%%%%%%%%%

Combining this with \eqref{eq:AtconsttypeD}, it follows that the (possible) zeros of the conformal level $L$ vacuum determinant in a fixed-root sector for $V_k(\sof_8)$ can be classified as,
\begin{equation} \label{eq: so8prelimzero}
    k = \begin{cases}
        k_{6,q} = -6 + \frac{6}{q} \ \ : q \leqslant L \ \ \text{and} \ \ (q,6)=1 \\ 
        k_{2,q} = -6 + \frac{2}{q} \ \ : q \leqslant \lfloor \frac{L}{3} \rfloor \ \ \text{and} \ \ (q,2)=1 \\
        k_{3,q} = -6 + \frac{3}{q} \ \ : q \leqslant \lfloor \frac{L}{7} \rfloor \ \ \text{and} \ \ (q,3)=1 \\
        k_{A,B} = -6 + \frac{A}{B}: A \geqslant 8 \ \ \text{and} \ \ B \leqslant L
    \end{cases}
\end{equation}
To implement graded unitarity constraints for $L \geqslant 13$, we first determine some of the smallest zeros of the vacuum determinant in the fixed-root sector. 

Consider $L = (1,2,0)~\textrm{mod}~6$. From \eqref{eq: so8prelimzero} and the data in Table \ref{tb:multandLevforso8}, it follows that for $L = 1~\textrm{mod}~6$, the most negative zero of the fixed root conformal level $L$ vacuum determinant is $k = k_{6,L}$, and the multiplicity of this zero is one. Furthermore, note that for $L \geqslant 13$,
\begin{equation} \label{eq:so8inequality}
    k_{6,L-2} < k_{8,L}~,
\end{equation}
which implies that the second most negative zero of the fixed-root vacuum determinant at conformal level $L$ is $k = k_{6,L-2}$. Similarly, for $L = (2,0)~\textrm{mod}~6$, the most negative zero is $k = k_{6,L-1}$. 

When $L = 3~\textrm{mod}~6$, from Table \ref{tb:multandLevforso8}, it follows that the most negative zero is $k = k_{6,L}$ with multiplicity two, and from \eqref{eq:so8inequality}, the second most negative zero is $k = k_{6,L-2}$. When $L = 4~\textrm{mod}~6$, similar inequality arguments imply that the most negative zero is $k = k_{6,L-1}$ with multiplicity 32, and the second most negative zero is $k = k_{6,L-3}$. Furthermore, for $L = 5~\textrm{mod}~6$, \eqref{eq: so8prelimzero} and the data in Table \ref{tb:multandLevforso8} imply that the most negative zero is $k = k_{6,L}$ with multiplicity one, and the second most negative zero is $k = k_{6,L-2}$ with multiplicity 322.\footnote{It is a technical, but an important point that at $L = 4~\textrm{mod}~6$ and $L = 5~\textrm{mod}~6$, there is no \emph{new} primitive null state at $k = k_{6,L-1}$ and $k = k_{6,L-2}$, respectively, and thus, the evenness of the multiplicity of these zeros follows from the multiplicity of two in the fixed root $L = 3~\textrm{mod}~6$ vacuum determinant. The precise multiplicities of 32 and 322 can be computed by expanding the $1/R$ factor in \eqref{eq:multAKMmain}.} As highlighted by the bespoke analysis, it is relevant to determine the third most negative zero of the vacuum determinant at $L=5~\text{mod}~6$ to implement determinant-based graded unitarity constraints in their full glory. To this end, note that for $L \geqslant 13$, and $L=5~\text{mod}~6$, 
\begin{equation}
    k_{6,L-4} < k_{8,L}~,
\end{equation}
which implies that the third most negative zero of the fixed-root vacuum determinant at $L \geqslant 13$ and $L=5~\text{mod}~6$ is $k_{6,L-4}$. 

To summarise, we obtain the following structure of zeros for the level $L \geqslant 13$ vacuum determinant in a fixed-root sector,
\begin{itemize}
    \item At level $L = 1~\text{mod}~6$, the most negative zero is given by $k=k_{6,L}$, and this zero always occurs with multiplicity one. Moreover, the second most negative zero is given by $k = k_{6,L-2}$. 
    \item At level $L = (0,2)~\text{mod}~6$ and $L$ even, the most negative zero is given by $k=k_{6,L-1}$.
    \item At level $L = 3~\text{mod}~6$, the most negative zero is given by $k = k_{6,L}$, and this zero always occurs with multiplicity two. Moreover, the second most negative zero is given by $k = k_{6,L-2}$. 
    \item At level $L = 4~\text{mod}~6$, the most negative zero is given by $k = k_{6,L-1}$ and this zero always occurs with multiplicity 32. Moreover, the second most negative zero is given by $k = k_{6,L-3}$. 
    \item At level $L = 5~\text{mod}~6$, the most negative zero is given by $k = k_{6,L}$, and this zero always occurs with multiplicity one. Moreover, the second most negative zero is given by $k = k_{6,L-2}$ and this zero always occurs with multiplicity 322. The third most negative zero is given by $k = k_{6,L-4}$.  
\end{itemize}

\subsubsection{Graded unitarity constraints on the level}
The conclusion of the structure of zeros established above is that the level $L \geqslant 13$ vacuum determinant in a fixed-root sector takes the following form,\\
\\
\underline{\textbf{Case 1: } $L = (1,2)~\text{mod}~6$}
\begin{equation} \label{eq:so8vacdet1}
\begin{split}
    &\text{det}~S_{L\delta+\alpha} \propto \textcolor{blue}{(k-k_{6,L})^1}(k-k_{6,L-2})^{\#_1}( \ldots \ldots) \qquad L = 1~\text{mod}~6~, \\
    &\text{det}~S_{L\delta+\alpha} \propto \textcolor{blue}{(k-k_{6,L-1})^{\#_2}}( \ldots \ldots) \qquad \qquad \qquad \ \ L = 2~\text{mod}~6~.
\end{split}
\end{equation}
\\
\\
\underline{\textbf{Case 2: } $L = (3,4,5,0)~\text{mod}~6$}
\begin{equation} \label{eq:so8vacdet2}
\begin{split}
    &\text{det}~S_{L\delta+\alpha} \propto \textcolor{blue}{(k-k_{6,L})^2}(k-k_{6,L-2})^{\#_1}( \ldots \ldots) \qquad \qquad \qquad \qquad \ L = 3~\text{mod}~6~, \\
    &\text{det}~S_{L\delta+\alpha} \propto \textcolor{blue}{(k-k_{6,L-1})^{32}}(k-k_{6,L-3})^{\#_2}( \ldots \ldots) \qquad \qquad \qquad \ \ \ L = 4~\text{mod}~6~, \\
    &\text{det}~S_{L\delta+\alpha} \propto \textcolor{blue}{(k-k_{6,L})^1(k-k_{6,L-2})^{322}}(k-k_{6,L-4})^{\#_3}( \ldots \ldots) \qquad L = 5~\text{mod}~6~, \\
    &\text{det}~S_{L\delta+\alpha} \propto \textcolor{blue}{(k-k_{6,L-1})^{\#_4}}( \ldots \ldots) \qquad \qquad \qquad \qquad \qquad \qquad \ \ \ L = 0~\text{mod}~6~.
\end{split}
\end{equation}

The all-level determinant based graded unitarity constraints now follow in a straightforward manner. In particular, note that when $L = 1~\text{mod}~6$, graded unitarity sign constraints rule out $k \in (k_{6,L},k_{6,L-2})$, whereas when $L = 2~\text{mod}~6$, there are no new constraints. Furthermore, at conformal level $L = (3,4)~\text{mod}~6$, while it is important to note that the most negative zero has changed to $k = k_{6,L}$ or $k = k_{6,L-1}$, there are no new constraints from graded unitarity due to the multiplicity of the most negative zero being even. 

When $L = 5~\text{mod}~6$, graded unitarity sign constraints rule out the possibility of $k \in (k_{6,L}, k_{6,L-2}) \bigcup (k_{6,L-2}, k_{6,L-4})$. On the other hand, when $L = 0~\text{mod}~6$, there are again no new constraints. 

When combined with the low level graded unitarity constraints in \eqref{eq:lowlevelso8fullconst}, we find that with the assumption of the standard $\mfR$-filtration on $V_k(\mathfrak{so}_8)$ affine Kac--Moody algebras, all-level (fixed root) determinant-based graded unitarity constraints rule out all levels other than the following,
\begin{equation} \label{eq:so8alllevel}
    k = \begin{cases}
        \textcolor{knownSCFTtext}{k_{6,q} = -6 + \frac{6}{q}: (6,q) = 1} \\
        \textcolor{knownSCFTtext}{-2} \\
        \textcolor{conjecturedSCFTtext}{-\frac{26}{5}} \\
        k_{2,q} = -6 + \frac{2}{q}: (2,q) = 1 \\
        -\frac{5}{2}
    \end{cases}
\end{equation}
The first set of levels above is the set of boundary admissible levels. The $V_k(\sof_8)$ affine Kac--Moody algebras corresponding to these levels are precisely the ones associated with the ``class-I" $D_q^6(\mathfrak{so}_8)$ Argyres--Douglas SCFTs under the SCFT/VOA correspondence \cite{Xie:2016evu, Song:2017oew}. The level $k = -2$ (while non-admissible) is precisely the one corresponding to the associated VOA $V_{-2}(\sof_8)$ of the Lagrangian $SU(2)$ with $N_f = 4$ flavours SCFT. 

While the level $-\frac{26}{5}$ is non-admissible, $V_{-\frac{26}{5}}(\mathfrak{so}_8)$ has been conjectured to be the associated VOA of the $D^4_5(\mathfrak{so}_8)$ Argyres--Douglas SCFT which has no extra mass parameters/flavour symmetries coming from the irregular puncture. Furthermore, this affine VOA has also been conjectured to be quasi-lisse \cite{adamovic2026new}. Nevertheless, as discussed in Section \ref{sec:intro}, the associated VOA of $D^4_5(\mathfrak{so}_8)$ need not coincide with the affine Kac--Moody algebra $V_{-\frac{26}{5}}(\mathfrak{so}_8)$, but may instead be a non-trivial finite extension thereof. On the other hand, the $V_k(\mathfrak{so}_8)$ affine VOAs corresponding to the remaining set of levels $k_{2,q}$ and  $-\frac{5}{2}$ are not currently known (or conjectured) to be associated with any unitary 4d $\mathcal{N} = 2$ SCFT. In either case, although these levels are not excluded by all-level (fixed root) determinant-based graded unitarity constraints, it remains to be seen whether analogous constraints from a different weight sector or finer state-based constraints exclude any of them. 

\subsection{$V_k(\sof_{2N})$}
We now turn to implementing graded unitarity constraints for $V_k(\sof_{2N})$ with $N$ even. The structure of zeros of the fixed-root vacuum determinant is a natural generalisation of the preceding discussion for $V_k(\mathfrak{so}_8)$.

\subsubsection{Low level graded unitarity constraints}
We begin by performing a low level analysis for conformal levels $ \leqslant 6$. To this end, note that the fixed-root vacuum determinant for $L = 1,2,3,4$ is given by,
\begin{equation} \label{eq: so2NevenL1to3}
\begin{split}
    &\text{det}~S_{\delta + \alpha} \ \propto \textcolor{blue}{(k - 0)}~, \\
    &\text{det}~S_{2\delta + \alpha} \propto \textcolor{blue}{(k - k_{N,1})}(k-k_{2N-2,1})^{\#_1} (\ldots \ldots)~,\\ 
    &\text{det}~S_{3\delta + \alpha} \propto \textcolor{blue}{(k - k_{2N-2,3})}(k-k_{N,1})^{\#_2} (\ldots \ldots)~,\\
    &\text{det}~S_{4\delta + \alpha} \propto \textcolor{blue}{(k - k_{N,2})}(k-k_{2N-2,3})^{\#_3} (\ldots \ldots)~.\\
\end{split}   
\end{equation}
Hence, up to $L = 4$, graded unitarity constraints require the level $k$ to be in the regime,
\begin{equation}
    -(2N-2) < k \leqslant k_{N,2} \qquad \text{or} \qquad k \in \{k_{2N-2,3}, k_{N,1}, 0\}~. 
\end{equation}
At conformal level $L = 5$, one obtains that the fixed-root vacuum determinant is given by,
\begin{equation} \label{eq:level5Dtypeeven}
\begin{split}
    &\text{det}~S_{5\delta + \alpha} \propto \textcolor{blue}{(k - k_{10,5})^2}(k-k_{6,2})^{\#_1} (\ldots \ldots) \qquad \ \ \ \text{when} \ \ N = 6~, \\
    &\text{det}~S_{5\delta + \alpha} \propto \textcolor{blue}{(k - k_{2N-2,5})}(k-k_{N,2})^{\#_2} (\ldots \ldots) \qquad \text{when} \ \ N \geqslant 8~.   
\end{split}
\end{equation}
The contrasting multiplicities of the most negative zero of the fixed root $L = 5$ determinant for $N = 6$ and $N \geqslant 8$ are a straightforward consequence of \eqref{eq:Lmultofk2Nm2}. Hence, there are no new constraints from graded unitarity for $N = 6$, whereas for $N \geqslant 8$, graded unitarity restricts the level $k$ to be in the regime, 
\begin{equation}
    -(2N-2) < k \leqslant k_{2N-2,5} \qquad \text{or} \qquad k \in \{k_{N,2}, k_{2N-2,3}, k_{N,1}, 0\}~. 
\end{equation}
At $L=6$, the fixed-root vacuum determinant is given by,
\begin{equation} \label{eq:Lvl6soeven}
\begin{split}
    &\text{det}~S_{6\delta + \alpha} \ \propto \textcolor{blue}{(k - k_{10,5})^{\mathrm{even}}(k - k_{8,3})^4}(k-k_{6,2})^{\#_1} (\ldots \ldots) \qquad \ \text{when} \ \ N = 6~, \\
    &\text{det}~S_{6\delta + \omega_4} \propto \textcolor{blue}{(k - k_{10,5})^{\mathrm{even}}(k - k_{8,3})^1}(k-k_{6,2})^{\#_2} (\ldots \ldots) \qquad \ \text{when} \ \ N = 6~, \\
    &\text{det}~S_{6\delta + \alpha} \ \propto \textcolor{blue}{(k - k_{N,3})}(k-k_{2N-2,5})^{\#_3} (\ldots \ldots) \qquad \qquad \qquad \ \ \ \text{when} \ \ N \geqslant 8~,   
\end{split}
\end{equation}
where the even multiplicity of $(k-k_{10,5})$ follows from similar arguments as $V_k(\mathfrak{so}_8)$: at $L = (N,N+1)~\mathrm{mod}~(2N-2)$, there is no new primitive null at $k = k_{2N-2,L-1}$ and $k = k_{2N-2,L-2}$, respectively, and the evenness then follows from the multiplicity of two at $L = (N-1)~\mathrm{mod}~(2N-2)$. Note that the multiplicity of four of the $k = k_{8,3}$ zero is a consequence of the null state being a component of the $[0,0,0,1,0,0]$ representation, which has root multiplicity four. Hence, when $N = 6$, there are no new constraints from graded unitarity in the fixed-root vacuum determinant sector. However, considering the vacuum determinant in the weight $\lambda = \omega_4$ sector instead,\footnote{It is worth mentioning that even in the weight $\omega_4$ sector, the multiplicity of $k=k_{6,2}$ is indeed non-zero. This follows from the fact that the primitive null at $k = k_{6,2}$ is a component of the $[0,0,0,0,0,2]$ representation of $\sof_{12}$; and the multiplicity of $\lambda = \omega_4$ in this representation is non-zero. The evenness of the $k = k_{10,5}$ multiplicity persists at $L = 6$ and follows from \eqref{eq:level5Dtypeeven}.} graded unitarity constraints for $N = 6$ eliminate the interval $k \in (k_{8,3}, k_{6,2})$.  

On the other hand, when $N \geqslant 8$, graded unitarity constraints imply,
\begin{equation}
    -(2N-2) < k \leqslant k_{N,3} \qquad \text{or} \qquad k \in \{k_{2N-2,5}, k_{N,2}, k_{2N-2,3}, k_{N,1}, 0\}~. 
\end{equation}
The preceding analysis reveals a systematic alternation between the exclusion of the levels $(k_{2N-2,L}, k_{2N,L-1})$ and $(k_{2N,L}, k_{2N-2,L-1})$. 

\subsubsection{Structure of zeros of the vacuum determinant} \label{sec:soevenstruczero}
Recall that the possible zeros in the conformal level $L$ vacuum determinant are given by \eqref{eq: so2nprelimzero}. The precise structure of zeros in the various regimes of the level $L$ then follows from certain inequalities. For instance, note that for $6 < L \leqslant N+1$ and $N \geqslant 10$,\footnote{When $N = 6$ and $N = 8$, the cases of $L \leqslant 4$ and $L \leqslant 6$ respectively, are already handled by bespoke analysis.} 
\begin{equation} \label{eq:soevenstr1}
    k_{2N, L-1} < k_{4N-8, L}~.
\end{equation}
This implies that for $6 < L \leqslant N-2$ and $N \geqslant 10$, one has the following structure of zeros,
\begin{itemize}
    \item For $L$ odd, the most negative zero of the fixed-root vacuum determinant is $k = k_{2N-2,L}$, and this zero always occurs with multiplicity one. Furthermore, the second most negative zero is $k = k_{2N,L-1}$.
    \item For $L$ even, the most negative zero of the fixed-root vacuum determinant is $k = k_{2N,L}$, and this zero always occurs with multiplicity one. Furthermore, the second most negative zero is $k = k_{2N-2,L-1}$.
\end{itemize}
Consequently, determinant-based constraints from graded unitarity up to $L = N-2$ restrict the level $k$ to be in the regime,
\begin{equation} \label{eq:constSOevenuptoNm2}
    -(2N-2) < k \leqslant k_{2N, N-2} \ \ \text{or} \ \ k \in \{k_{2N-2, N-3}, k_{2N, N-4}, \ldots, k_{2N-2, 3}, k_{2N, 2}, k_{2N-2,1}\}~.
\end{equation}
As a consequence of the subtle nature of the multiplicities of zeros when $L = (N-1)~\text{mod}~(2N-2)$, we will perform a separate analysis for the structure of zeros when $L = (N-1,N,N+1)~\text{mod}~(2N-2)$. To begin with, note that the structure of zeros at $L = N-1, N, N+1$ for $N \geqslant 6$ is given by the following.
\begin{itemize}
    \item At $L = N-1$, the most negative zero of the fixed-root vacuum determinant is given by $k = k_{2N-2, N-1} = k_{2,1}$, and this zero always occurs with multiplicity two. Furthermore, the second most negative zero is $k = k_{2N, N-2}$. 
    \item At $L = N$, 
    \begin{enumerate}
        \item When $N = 6$, the structure of zeros has already been discussed in \eqref{eq:Lvl6soeven}.
        \item On the other hand, for $N \geqslant 8$, the most negative zero of the fixed-root vacuum determinant is given by $k = k_{2,1}$, and this zero always occurs with even multiplicity. Furthermore, the second most negative zero is given by $k = k_{2N, N-2}$. 
    \end{enumerate}
    \item At $L = N+1$, the most negative zero of the fixed-root vacuum determinant is given by $k = k_{2N-2,N+1}$, and this zero always occurs with multiplicity one. 
    \begin{enumerate}
        \item When $N = 6$, the second most negative zero is given by $k = k_{2,1}$, and this zero occurs with even multiplicity. Furthermore, the third most negative zero for $N = 6$ is given by $k = k_{8,3}$. 
        \item On the other hand, for $N \geqslant 8$, the second most negative zero is given by $k = k_{2N-2, N-1} = k_{2,1}$, and this zero occurs with even multiplicity. Moreover, the third most negative zero is given by $k = k_{2N, N-2}$.   
    \end{enumerate}
\end{itemize}
From the above structure of zeros, it follows that up to $L = N+1$, determinant-based graded unitarity constraints rule out all levels $k$ except,
\begin{equation} \label{eq:so4nlevellessNp2}
\begin{split}
    &-10 < k \leqslant k_{10,7} \ \ \text{or} \ \ k \in \{k_{10,i}: i \in \mathcal{I}_6; k_{12,j}: j \in \mathcal{J}_6\} \ \ \text{or} \ \ k = k_{8,3} \qquad \qquad \text{for} \ \ N=6~, \\
    &-(2N-2) < k \leqslant k_{2N-2,N+1} \ \ \text{or} \ \ k \in \{k_{2N-2,i}: i \in \mathcal{I}_N, k_{2N,j}: j \in \mathcal{J}_N \} \qquad \text{for} \ \ N \geqslant 8~, \\
\end{split}
\end{equation}
where $\mathcal{I}_N$ and $\mathcal{J}_N$ are the sets,
\begin{equation}
\begin{split}
    &\mathcal{I}_N = \{1, 3, 5, \ldots, N-1\}~, \\ 
    &\mathcal{J}_N = \{2, 4, 6, \ldots, N-2\}~.
\end{split}
\end{equation}
Now we proceed to describing the structure of zeros for $L \geqslant N+2$, ultimately leading to all-level determinant-based graded unitarity constraints. 

Note that for $N \geqslant 5$ and $L \geqslant N+2$,
\begin{equation}
    k_{2N-2,L-2} < \{k_{2N,L-1}, \ k_{4N-8, L}\}, \qquad k_{2N-2,L-1} < \{k_{2N,L}, \ k_{4N-8, L}\}~,
\end{equation}
and one obtains the following structure of zeros for $L \geqslant N+2$ and $L \neq (N-1,N,N+1)~\text{mod}~(2N-2)$, 
\begin{itemize}
    \item For $L$ odd, the most negative zero is given by $k = k_{2N-2, L}$, and this zero always occurs with multiplicity one. Furthermore, the second most negative zero is given by $k = k_{2N-2, L-2}$. 
    \item For $L$ even, the most negative zero is given by $k = k_{2N-2,L-1}$. 
\end{itemize}
As elaborated previously in this section, having more refined knowledge of the structure of zeros for $L = (N-1,N,N+1)~\text{mod}~(2N-2)$ is desirable. To this end, we note that for $N \geqslant 5$ and $L \geqslant N+2$, the structure of zeros in these cases is given by,
\begin{itemize}
    \item At $L = (N-1)~\text{mod}~(2N-2)$, the most negative zero of the fixed-root vacuum determinant is given by $k = k_{2N-2, L}$, and this zero always occurs with multiplicity two. Furthermore, the second most negative zero is given by $k = k_{2N-2, L-2}$. 
    \item At $L = N~\text{mod}~(2N-2)$, the most negative zero of the fixed-root vacuum determinant is given by $k = k_{2N-2, L-1}$, and this zero always occurs with even multiplicity. Furthermore, one obtains: 
    \begin{enumerate}
        \item For $L = 3N-2$, the second most negative zero is given by $k = k_{2N, 3N-2}$, and this zero occurs with multiplicity one. Furthermore, the third most negative zero is given by $k = k_{2N-2, 3N-5}$.
        \item For $L \neq 3N-2$, the second most negative zero is given by $k = k_{2N-2, L-3}$. 
    \end{enumerate}
    \item At $L = (N+1)~\text{mod}~(2N-2)$, the most negative zero of the fixed-root vacuum determinant is given by $k = k_{2N-2, L}$, and this zero always occurs with multiplicity one. The second most negative zero is given by $k = k_{2N-2,L-2}$, and this zero always occurs with even multiplicity. Furthermore, one obtains: 
    \begin{enumerate}
        \item For $L = 3N-1$, the third most negative zero is given by $k = k_{2N, 3N-2}$.
        \item For $L \neq 3N-1$, the third most negative zero is given by $k = k_{2N-2, L-4}$. 
    \end{enumerate}
\end{itemize}

\subsubsection{Graded unitarity constraints for $V_k(\sof_{2N})$ with $N$ even}
To summarise the structure of zeros of the previous section, one obtains that the fixed-root vacuum determinant for $N \geqslant 6$ and $L \geqslant N+2$ takes the following form, \\
\\
\underline{\textbf{Case 1: } $L = (1,2,\ldots, N-2, N+2, \ldots, 2N-2)~\text{mod}~(2N-2)$}
\begin{equation}
\begin{split}
    &\text{det}~S_{L\delta+\alpha} \propto \textcolor{blue}{(k-k_{2N-2,L})^1}(k-k_{2N-2,L-2})^{\#}( \ldots \ldots) \qquad L \ \ \text{odd}~, \\
    &\text{det}~S_{L\delta+\alpha} \propto \textcolor{blue}{(k-k_{2N-2,L-1})^{\#}}( \ldots \ldots) \qquad \qquad \qquad \qquad \ L \ \ \text{even}~.
\end{split}
\end{equation}

\underline{\textbf{Case 2: } $L = (N-1)~\text{mod}~(2N-2)$}
\begin{equation}
\begin{split}
    &\text{det}~S_{L\delta+\alpha} \propto \textcolor{blue}{(k-k_{2N-2,L})^2}(k-k_{2N-2,L-2})^{\#}( \ldots \ldots)~.
\end{split}
\end{equation}

\underline{\textbf{Case 3: } $L = N~\text{mod}~(2N-2)$}
\begin{equation}
\begin{split}
    &\text{det}~S_{L\delta+\alpha} \propto \textcolor{blue}{(k-k_{2N-2,3N-3})^{\mathrm{even}}(k-k_{2N,3N-2})^1}(k-k_{2N-2, 3N-5})^{\#}( \ldots) \qquad L=3N-2~, \\
    &\text{det}~S_{L\delta+\alpha} \propto \textcolor{blue}{(k-k_{2N-2,L-1})^{\mathrm{even}}}(k - k_{2N-2, L-3})^{\#}( \ldots) \qquad \qquad \qquad \qquad \ \ \ \ \ L \neq 3N-2~.
\end{split}
\end{equation}

\underline{\textbf{Case 4: } $L = (N+1)~\text{mod}~(2N-2)$}
\begin{equation}
\begin{split}
    &\text{det}~S_{L\delta+\alpha} \propto \textcolor{blue}{(k-k_{2N-2,3N-1})^{1}(k-k_{2N-2,3N-3})^{\mathrm{even}}}(k-k_{2N, 3N-2})^{\#}(\ldots) \qquad L=3N-1~, \\
    &\text{det}~S_{L\delta+\alpha} \propto \textcolor{blue}{(k-k_{2N-2,L})^{1} (k - k_{2N-2, L-2})^{\mathrm{even}}}(k - k_{2N-2, L-4})^{\#}(\ldots) \qquad \ \ \ \ L \neq 3N-1~.
\end{split}
\end{equation}

This structure of the fixed-root vacuum determinant makes the implementation of graded unitarity constraints immediate. In particular, note that for $L \geqslant N+2$ and $L = \{1,2,\ldots, N-2, N+2, \ldots, 2N-2\}~\text{mod}~(2N-2)$, determinant-based graded unitarity sign constraints eliminate the region $k \in (k_{2N-2, L}, k_{2N-2, L-2})$ for $L$ odd, whereas, they do not give any constraints for $L$ even. Similarly, for $L = (N-1)~\text{mod}~(2N-2)$ (including $L = 3N-3$), there are no new constraints from graded unitarity. When $L = N~\text{mod}~(2N-2)$, graded unitarity eliminates $k \in (k_{2N, 3N-2}, k_{2N-2, 3N-5})$ for $L = 3N-2$, whereas it imposes no constraints for $L \neq 3N-2$. When $L = (N+1)~\text{mod}~(2N-2)$, graded unitarity constraints further rule out $k \in (k_{2N-2, 3N-1}, k_{2N-2, 3N-3}) \bigcup (k_{2N-2, 3N-3}, k_{2N, 3N-2})$ at $L = 3N-1$, and $k \in (k_{2N-2, L}, k_{2N-2, L-2}) \bigcup (k_{2N-2, L-2}, k_{2N-2, L-4})$ at $L \neq 3N-1$. Combining this with the graded unitarity constraints up to $L = N+1$, \emph{i.e.}, \eqref{eq:so4nlevellessNp2}, one obtains that fixed root determinant-based graded unitarity rules out all levels $k$ except,
\begin{equation} \label{eq:VkSOevengrunitresults}
\begin{split}
    &N = 6: \qquad k = \begin{cases}
        -10 + \frac{10}{q}: (10,q) = 1 \\ 
        -4, -7, -\frac{37}{4} \\
        -10 + \frac{2}{q}:  (2,q) = 1 \\
        -\frac{22}{3}
    \end{cases} \\
    &N \geqslant 8: \qquad k = \begin{cases}
        -(2N-2) + \frac{(2N-2)}{q}: (2N-2, q) = 1 \\
        -(2N-2) + \frac{(2N-2)}{q}: q \in 2\mathbb{N}+1 \ \ \text{and} \ \ (2N-2, q) \neq 1 \\
        -(2N-2) + \frac{N}{q} \qquad : q \in \mathbb{N} \ \ \text{and} \ \ 1 \leqslant q \leqslant \frac{N-2}{2} \\
        -(2N-2) + \frac{2N}{3N-2}
    \end{cases}
\end{split}
\end{equation}
The analysis for $V_k(\mathfrak{so}_{2N})$ with $N$ odd proceeds analogously, although the structure of zeros of the vacuum determinant differs due to the contrasting multiplicities in \eqref{eq:Lmultofk2Nm2}. We defer the details to Appendix \ref{app:VkSOodd}, and combine the resulting restrictions on the level $k$ with those obtained above for even $N$.

We find that with the assumption of the standard $\mfR$-filtration \eqref{eq:AKMRfiltration} on $V_k(\mathfrak{so}_{2N})$ for $N \geqslant 5$, fixed root determinant-based graded unitarity rules out all levels $k$ except,  
\begin{equation} \label{eq:VkSOfullresults}
    k = \begin{cases}
        \textcolor{knownSCFTtext}{-(2N-2) + \frac{(2N-2)}{q}: (2N-2, q) = 1} \\
        \textcolor{conjecturedSCFTtext}{-(2N-2) + \frac{(2N-2)}{q}: q \in 2\mathbb{N}+1 \ \ \text{and} \ \ (2N-2, q) \neq 1} \\
        \textcolor{conjecturedSCFTtext}{-(2N-2) + \frac{N}{q} \qquad : 1 \leqslant q \leqslant \frac{N-2}{2}  \ \ \text{and} \ \  \frac{N}{\text{gcd}(N,q)} \in 2\mathbb{N}} \\
        \textcolor{conjecturedSCFTtext}{-(2N-2) + \frac{2N}{3N-2} \ \ : N \in 4\mathbb{N}} \\
        -(2N-2) + \frac{N}{q} \qquad : 1 \leqslant q \leqslant \frac{N-2}{2} \ \ \text{and} \ \ \frac{N}{\text{gcd}(N,q)} \in 2\mathbb{N}+1  \\
        -(2N-2) + \frac{2N}{3N-2} \ \ : N \in 4\mathbb{N}+2 \\
        -(2N-2) + \frac{2}{q} \qquad \ : q \in 2\mathbb{N}+1 \ \ \text{and} \ \ N \in 2\mathbb{N}+1 \\
        -\frac{22}{3} \qquad \qquad \qquad \ \ \ :N = 6
    \end{cases}
\end{equation}
The first set of levels above is the set of boundary admissible levels, and the $V_k(\sof_{2N})$ affine Kac--Moody algebras corresponding to these levels are precisely the ones associated with the ``class-I" $D_q^{2N-2}(\mathfrak{so}_{2N})$ Argyres--Douglas SCFTs under the SCFT/VOA correspondence \cite{Xie:2016evu, Song:2017oew}. 

While the second, third, and fourth sets of levels represented in green are all non-admissible, the corresponding affine VOAs are conjectured to be associated with the $D_q^{2N-2}(\mathfrak{so}_{2N})$, $D_q^{N}(\mathfrak{so}_{2N})$, and $D_{\frac{3N-2}{2}}^{N}(\mathfrak{so}_{2N})$ Argyres--Douglas SCFTs, respectively.\footnote{It is worth mentioning that while the $V_{-N+2}(\mathfrak{so}_{2N})$ affine Kac--Moody algebras are non-admissible, they are known to be quasi-lisse \cite{arakawa2017sheets}.} As alluded to in Section \ref{sec:intro}, the non-admissibility of these levels means that the associated VOAs of these theories could be non-trivial finite extensions of the corresponding affine Kac--Moody algebras. On the other hand, the remaining set of non-admissible $V_k(\mathfrak{so}_{2N})$ are not currently known to be associated with any unitary 4d $\mathcal{N} = 2$ SCFTs under the SCFT/VOA correspondence. In either case, we leave open the possibility that an analogous analysis in a different weight sector or a finer state-based graded unitarity analysis might rule out some or all of these non-admissible levels.

%% file: sections/sec_6_exceptional_gr_unit.tex
%%%%%%%%%
\section{Graded unitarity constraints on $V_k(\mathfrak{e}_{n=6,7,8})$} \label{sec:typeEgrunit}
%%%%%%%%%
In this section, we complete our characterisation by studying graded unitarity constraints on simply-laced exceptional affine Kac--Moody algebras. The approach will be similar to that used in Sections \ref{sec:typeAgrunit} and \ref{sec:typeDgrunit}. We will first perform a low-level graded unitarity analysis, assemble a variety of data about the representations (and their root multiplicities) of the primitive null states corresponding to a specific set of zeros of the vacuum determinant, and then impose all-level determinant-based graded unitarity sign constraints from \eqref{eq:grunitsignrule}, leading to constraints on the level $k$.\footnote{All calculations in this section were performed using \texttt{SageMath}. For all relevant cases, the primitive null representations first appeared in \cite{jiang2025singular}, while for many cases, the relevant root multiplicities can be found in \cite{Lübeck_2001}.} While we investigate the $V_k(\mathfrak{e}_6)$ case in necessary detail, we keep the discussion of the $V_k(\mathfrak{e}_7)$ and $V_k(\mathfrak{e}_8)$ affine VOAs brief and mention only the relevant results of various computations.

%%%%%%%%%
\subsection{$V_k(\mathfrak{e}_6)$}
%%%%%%%%%
\input{E6Dynkin}
%%%%%%%%%
For the $V_k(\mathfrak{e}_6)$ affine VOA, we follow the usual convention in which the Dynkin diagram of $E_6$ is given by Figure \ref{fig:E6Dynkindiagram}. Before investigating graded unitarity constraints, as with the analyses in the previous sections, we first note that the constraints on $A_{(t,\alpha)}$ that are consistent with the regularity criterion \eqref{eq:ADEregularity} are given by,
\begin{equation} \label{eq:Atconste6}
    A_{(t,\alpha)} \ \ \begin{cases}
        = 12 \qquad \text{when} \ \ t = 1 \ \ \text{and} \ \ \alpha = \theta \\ = 12 \qquad \text{when} \ \ t = 2 \ \ \text{and} \ \ \alpha = (0,1,1,1,1,0) =: \alpha^{\mathfrak{e}_6}_{(2)} \\ = 12 \qquad \text{when} \ \ t = 3 \ \ \text{and} \ \ \alpha = (0,0,0,1,0,0) =: \alpha^{\mathfrak{e}_6}_{(3)} \\ \geqslant 18 \qquad \text{otherwise}
    \end{cases}
\end{equation}
Moreover, note that while $2\alpha^{\mathfrak{e}_6}_{(2)}$ and $3\alpha^{\mathfrak{e}_6}_{(3)}$ are not regular dominant, their regular dominant representatives in the Weyl orbit can be found using a series of Weyl reflections to be the highest root. As a result, one obtains that, 
\begin{equation}
    \frac{E(\theta)}{E(0)} = -\frac{E(2\alpha^{\mathfrak{e}_6}_{(2)})}{E(0)} = -\frac{E(3\alpha^{\mathfrak{e}_6}_{(3)})}{E(0)} = \text{ch Ad}~.
\end{equation}
%%%%%%%%%%%
\subsubsection{Bespoke low-level analysis}
In this case, it is useful to perform a low-level analysis up to $L = 4$. To this end, the fixed-root vacuum determinant for conformal levels $L \leqslant 4$ takes the following form, 
\begin{equation} \label{eq:lowlevele6det1to4}
\begin{split}
    &\text{det}~S_{\delta + \alpha} \qquad \propto \textcolor{blue}{(k - 0)}~, \\
    &\text{det}~S_{2\delta + \alpha} \ \ \ \ \ \propto \textcolor{blue}{(k + 3)^5}(k - 0)^{\#_1} (\ldots \ldots)~, \\ 
    &\text{det}~S_{3\delta + \omega_1 + \omega_6} \propto \textcolor{blue}{(k + 6)(k + 4)^4} (k + 3 )^{\#} (\ldots \ldots)~, \\
    &\text{det}~S_{4\delta + \alpha} \ \ \ \ \ \propto \textcolor{blue}{(k + 9)^{1}(k + 15/2)^5}(k+6)^{\#} (\ldots \ldots)~. \\
\end{split}
\end{equation}
Up to $L  = 2$, determinant based graded unitarity constraints lead to $-12 < k \leqslant -3$ or $k = 0$. At $L = 3$, the null states at $k = -6$ and $k = -4$ are components of the $V_{\omega_1+\omega_6}$ and $V_{\omega_4}$ representations respectively. While determinant-based constraints already rule out $k \in (-6,-4) \bigcup (-4,-3)$, since the null states at $k = -6$ and $k = -4$ are components of different representations, passing to a representation basis further rules out $k = -4$. This leads to the constraints $-12 < k \leqslant -6$ or $k = \{-3,0\}$. At $L = 4$, the null states corresponding to the zeros at $k = -9$ and $k = -15/2$ are components of the adjoint and $V_{\omega_1+\omega_6}$ representation respectively, and a similar analysis as at $L = 3$ rules out $k \in (-9,-6)$. This leads to the following graded unitarity constraints up to $L \leqslant 4$,
\begin{equation} \label{eq:e6lowlevelconst}
    -12<k \leqslant -9 \qquad \text{or} \qquad k = \left\{-6, -3, 0 \right\}~.
\end{equation}
We now turn to investigating the structure of zeros for $L \geqslant 6$, and implementing all-level determinant-based graded unitarity constraints on $V_k(\mathfrak{e}_6)$.
%%%%%%%%%%%%%%%%%%%%%%%%%%%
\begin{table}[] 
    \centering
    \begin{tabular}{|c|c|c|c|}
        \hline\hline
        $k$ & $L_s$ & Representation & Multiplicity \\
        \hline\hline
        $k_{2,q}$  & $12q$ & $V_{\omega_4} = \textbf{2925}$ & $15$ \\
        $k_{3,q}$  & $4q$  & $V_\theta = \textbf{78}$ & $1$ \\
        $k_{4,q}$  & $12q$ & $V_{\omega_1+2\omega_2 + \omega_6} = \textbf{812175}$ & $1085$ \\
        $k_{6,q}$  & $3q$  & $V_{\omega_1 + \omega_6} = \textbf{650}$ & $5$ \\
        $k_{12,q}$ & $q$   & $V_\theta = \textbf{78}$ & $1$ \\
        \hline
    \end{tabular}
    \caption{\label{tb:e6zerodata} Smallest conformal level $L_s$, primitive null representations and their root multiplicities for different values of $k$.}
\end{table}
%%%%%%%%%%%%%%%%%%%%%%%%%%%
%%%%%%%%%%%%
\subsubsection{Structure of zeros and all-level graded unitarity constraints}
%%%%%%%%%%%%
We first assemble data relevant for understanding the structure of zeros of the fixed-root vacuum determinant. In particular, Table \ref{tb:e6zerodata} collects the smallest conformal level $L_s$ at which $k = k_{2,q}, k_{3,q}, k_{4,q}, k_{6,q}$, and $k_{12,q}$ first appear as zeros of the fixed-root vacuum determinant. In addition, we also note the representation of the primitive null state at these zeros, and the root multiplicity of the representation. 

To provide a glimpse into the computation of the representation and root multiplicity appearing in Table \ref{tb:e6zerodata}, note that a direct computation using \eqref{eq:expinchar} and \eqref{eq:Atconste6} implies,
\begin{equation} \label{eq:multrelevante6}
\begin{split}
    &\frac{M_{12/q,q}}{E(0)} = \frac{E(\theta)}{E(0)} = \text{ch}~\text{Ad} \qquad \qquad \qquad \qquad \qquad \qquad \text{when} \ \ (12,q) = 1~, \\
    &\frac{M_{3/q,4q}}{E(0)} = \frac{E(\theta)}{E(0)} + \frac{E(2\alpha^{\mathfrak{e}_6}_{(2)})}{E(0)} -\frac{E(3\alpha^{\mathfrak{e}_6}_{(3)})}{E(0)}= \text{ch}~\text{Ad} \qquad \text{when} \ \ (3,q) = 1~.
\end{split}
\end{equation}
As a result, the null states corresponding to the $k = k_{12,q}$ zero (with $(12,q) = 1$) at conformal level $L = q$, and the $k = k_{3,q}$ zero (with $(3,q) = 1$) at conformal level $L = 4q$ both are components of the adjoint representation of $\mathfrak{e}_6$. Consequently, in a fixed-root sector at the respective conformal levels $L$, the multiplicity of these zeros is one.

The relevant data collected in Table \ref{tb:e6zerodata}, together with \eqref{eq:Atconste6} implies that at conformal level $L$, the possible zeros of the fixed-root vacuum determinant are given by,
\begin{equation} \label{eq: e6prelimzero}
    k = \begin{cases}
        k_{12,q} = -12 + \frac{12}{q}: q \leqslant L \ \ \text{and} \ \ (q,12)=1 \\ 
        k_{2,q} = -12 + \frac{2}{q} \ \ : q \leqslant \lfloor \frac{L}{12} \rfloor \ \ \text{and} \ \ (q,2)=1 \\
        k_{3,q} = -12 + \frac{3}{q} \ \ : q \leqslant \lfloor \frac{L}{4} \rfloor \ \ \text{and} \ \ (q,3)=1 \\
        k_{4,q} = -12 + \frac{4}{q} \ \ : q \leqslant \lfloor \frac{L}{12} \rfloor \ \ \text{and} \ \ (q,4)=1 \\
        k_{6,q} = -12 + \frac{6}{q} \ \ : q \leqslant \lfloor \frac{L}{3} \rfloor \ \ \text{and} \ \ (q,6)=1 \\
        k_{A,B} = -12 + \frac{A}{B}: A \geqslant 18 \ \ \text{and} \ \ B \leqslant L
    \end{cases}
\end{equation}

It is convenient to consider the following set of conformal levels $\mathcal{L} = \{1,4,5,7,8,11\}~\text{mod}~12$\footnote{This set comprises all positive integers that are coprime to 12, and positive integers that are multiples of four that are coprime to three.} and split the analysis for conformal levels $L \in \mathcal{L}$ and $L \notin \mathcal{L}$. For a given conformal level $L \in \mathcal{L}$, it can be shown that for $L \geqslant 5$, \footnote{Note that $k = k_{12,L^-}$ is either of the form $k_{12,q}$ with $(12,q) = 1$ or of the form $k_{3,q}$ with $(3,q) = 1$.}
\begin{equation}
    k_{12,L^-} \leqslant k_{18,L}~,
\end{equation}
where $L^-$ denotes the next smallest conformal level in $\mathcal{L}$. As a consequence, \eqref{eq: e6prelimzero} and the multiplicities documented in Table \ref{tb:e6zerodata} lead to the following structure of zeros for the level $L \geqslant 5$ vacuum determinant in a fixed-root sector.
\begin{itemize}
    \item For $L \in \mathcal{L}$, the most negative zero is given by $k = k_{12,L}$ and this zero always appears with multiplicity one. Furthermore, the second most negative zero is given by $k = k_{12,L^-}$. 
    \item For $L \notin \mathcal{L}$, the most negative zero is given by $k = k_{12, L^-}$.
\end{itemize}

The all-level determinant-based graded unitarity constraints now follow almost immediately, given the previously established low-level constraints \eqref{eq:e6lowlevelconst}. First, note that at $L \geqslant 5$ and $L \in \mathcal{L}$, graded unitarity constraints rule out the region $k \in (k_{12, L}, k_{12, L^-})$. On the other hand, at $L \geqslant 5$ and $L \notin \mathcal{L}$, there are no new constraints from graded unitarity. Therefore, one obtains that all-level (fixed root) determinant-based graded unitarity constraints on $V_k(\mathfrak{e}_6)$ rule out all values of the level $k$, except the following.
\begin{equation} \label{eq:e6alllevel}
    k = \begin{cases}
        \textcolor{knownSCFTtext}{k_{12,q} = -12 + \frac{12}{q}: (12,q) = 1} \\
        \textcolor{conjecturedSCFTtext}{k_{3,q} = -12 + \frac{3}{q} \ \ : (3,q) = 1} \\
        \textcolor{knownSCFTtext}{-3} \\ 
        \textcolor{conjecturedSCFTtext}{-6}
    \end{cases}
\end{equation}
The $V_k(\mathfrak{e}_6)$ affine VOAs corresponding to the first set of levels $k = -12+\frac{12}{q}$ with $(12,q) = 1$ are \emph{boundary} admissible, and are precisely the ones associated to ``class-I'' $D_q^{12}(\mathfrak{e}_6)$ Argyres--Douglas SCFTs under the SCFT/VOA correspondence. Moreover, $V_{-3}(\mathfrak{e}_6)$ is precisely the associated VOA of the $E_6$ Minahan-Nemeschansky rank-one SCFT \cite{Minahan:1996fg}.

On the other hand, the affine VOAs corresponding to the levels $k = k_{3,q}$ and $k = -6$ are non-admissible and conjectured to be associated with the $D_{4q}^{12}(\mathfrak{e}_6)$ and $D_{2}^{12}(\mathfrak{e}_6)$ Argyres--Douglas SCFTs respectively. Being non-admissible, it could be that the associated VOAs of these theories are instead non-trivial finite extensions of affine Kac--Moody algebras. Thus, we leave open the possibility that an analogous analysis in a different weight sector or a finer, state-based graded unitarity analysis may rule out some or all of these non-admissible levels. 

%%%%%%%%%
\subsection{$V_k(\mathfrak{e}_7)$}
%%%%%%%%%
%%%%%%%%%
\input{E7Dynkin}
%%%%%%%%%
As with the case of $V_{k}(\mathfrak{e}_6)$, we find the following constraints on $A_{(t,\alpha)}$ that are consistent with the regularity criterion \eqref{eq:ADEregularity},
\begin{equation} \label{eq:Atconste7}
    A_{(t,\alpha)} \ \ \begin{cases}
        = 18 \qquad \text{when} \ \ t = 1 \ \ \text{and} \ \ \alpha = \theta \\ = 18 \qquad \text{when} \ \ t = 2 \ \ \text{and} \ \ \alpha = (1,1,1,2,1,1,0) =: \alpha^{\mathfrak{e}_7}_{(2)} \\ = 18 \qquad \text{when} \ \ t = 3 \ \ \text{and} \ \ \alpha = (0,0,1,1,1,0,0) =: \alpha^{\mathfrak{e}_7}_{(3)} \\ \geqslant 28 \qquad \text{otherwise}
    \end{cases}
\end{equation}
We again emphasise that while $2\alpha^{\mathfrak{e}_7}_{(2)}$ and $3\alpha^{\mathfrak{e}_7}_{(3)}$ are not regular dominant, their regular dominant representative in the Weyl orbit is the highest root. Consequently, one obtains that, 
\begin{equation}
    \frac{E(\theta)}{E(0)} = -\frac{E(2\alpha^{\mathfrak{e}_7}_{(2)})}{E(0)} = -\frac{E(3\alpha^{\mathfrak{e}_7}_{(3)})}{E(0)} = \text{ch Ad}~.
\end{equation}

In this case, the vacuum determinant in an appropriate fixed-weight sector for conformal levels $L \leqslant 5$ is given by, 
\begin{equation} \label{eq:lowlevele7det1to5}
\begin{split}
    &\text{det}~S_{\delta + \alpha} \ \ \ \propto \textcolor{blue}{(k - 0)}~, \\
    &\text{det}~S_{2\delta + \alpha} \ \ \propto \textcolor{blue}{(k + 4)^1}(k - 0)^{\#} (\ldots \ldots)~, \\ 
    &\text{det}~S_{3\delta + 2\omega_7} \propto \textcolor{blue}{(k + 8)} (k + 4 )^{\#} (\ldots \ldots)~, \\
    &\text{det}~S_{4\delta + \omega_6} \ \propto \textcolor{blue}{(k + 11)^{1}(k + 17/2)^4}(k+8)^{\#} (\ldots \ldots)~, \\
     &\text{det}~S_{5\delta + \alpha} \ \ \propto \textcolor{blue}{(k + 72/5)^{1}(k + 12)^5}(k+11)^{\#} (\ldots \ldots)~.
\end{split}
\end{equation}
Up to $L = 3$, determinant based graded unitarity constraints lead to $-18 < k \leqslant -8$ or $k = \{-4,0\}$. At $L = 4$, the null states corresponding to the zeros at $k = -11$ and $k = -17/2$ are components of different representations of $\mathfrak{e}_7$: the $V_{\omega_6}$ and $V_{2\omega_1}$ representation respectively. Thus, passing to a representation basis, graded unitarity rules out the entire region $k \in (-11,-8)$. Similarly, at $L = 5$, the null states corresponding to the zeros at $k = -72/5$ and $k = -12$ are components of the adjoint and $V_{2\omega_7}$ representation, implying that graded unitarity rules out the region $k \in (-72/5, -11)$. To summarise, graded unitarity constraints up to $L \leqslant 5$ restrict the level $k$ to be,
\begin{equation} \label{eq:e7lowlevelconst}
    k \leqslant -\frac{72}{5} \qquad \text{or} \qquad k = \left\{-11, -8, -4, 0 \right\}~.
\end{equation}

%%%%%%%%%%%%%%%%%%%%%%%
\begin{table}[]
    \centering
    \begin{tabular}{|c|c|c|c|}
        \hline\hline
        $k$ & $L_s$ & Representation & Multiplicity \\
        \hline\hline
        $k_{2,q}$  & $9q$ & $V_\theta =\textbf{133}$ & $1$ \\
        $k_{3,q}$  & $14q$  & $V_{\omega_2+\omega_7} = \textbf{40766}$ & $75$ \\
        $k_{6,q}$  & $5q$ & $V_{2\omega_7} = \textbf{1463}$ & $5$ \\
        $k_{9,q}$  & $6q$  & $V_{\omega_4} = \textbf{365750}$ & $390$ \\
        $k_{18,q}$ & $q$   & $V_\theta = \textbf{133}$ & $1$ \\
        \hline
    \end{tabular}
    \caption{\label{tb:e7zerodata} Smallest conformal level $L_s$, primitive null representations and their root multiplicities for different values of $k$.}
\end{table}
%%%%%%%%%%%%%%%%%%%%%%

From the relevant data assembled in Table \ref{tb:e7zerodata} and \eqref{eq:Atconste7}, we find that at conformal level $L$, the possible zeros in a fixed-root vacuum determinant are given by,
\begin{equation} \label{eq: e7prelimzero}
    k = \begin{cases}
        k_{18,q} = -18 + \frac{18}{q}: q \leqslant L \ \ \text{and} \ \ (q,18)=1 \\ 
        k_{2,q} = -18 + \frac{2}{q} \ \ : q \leqslant \lfloor \frac{L}{9} \rfloor \ \ \text{and} \ \ (q,2)=1 \\
        k_{3,q} = -18 + \frac{3}{q} \ \ : q \leqslant \lfloor \frac{L}{14} \rfloor \ \ \text{and} \ \ (q,3)=1 \\
        k_{6,q} = -18 + \frac{6}{q} \ \ : q \leqslant \lfloor \frac{L}{5} \rfloor \ \ \text{and} \ \ (q,6)=1 \\
        k_{9,q} = -18 + \frac{9}{q} \ \ : q \leqslant \lfloor \frac{L}{6} \rfloor \ \ \text{and} \ \ (q,9)=1 \\
        k_{A,B} = -18 + \frac{A}{B}: A \geqslant 28 \ \ \text{and} \ \ B \leqslant L
    \end{cases}
\end{equation}
As with $V_k(\mathfrak{e}_6)$, it is convenient to consider the following set of conformal levels $\mathcal{L} = \{1,5,7,9,11,13,17\}~\text{mod}~18$, and divide the analysis for $L \in \mathcal{L}$ and $L \notin \mathcal{L}$. \footnote{This set comprises all positive integers that are coprime to 18, and positive integers that are \emph{odd} multiples of nine.}  We find that the multiplicities documented in Table \ref{tb:e7zerodata}, and \eqref{eq: e7prelimzero} lead to the following structure of zeros for the level $L \geqslant 6$ vacuum determinant in a fixed-root sector.
\begin{itemize}
    \item For $L \in \mathcal{L}$, the most negative zero is given by $k = k_{18,L}$ and this zero always appears with multiplicity one. Furthermore, the second most negative zero is given by $k = k_{18,L^-}$, where $L^-$ denotes the next smallest conformal level in $\mathcal{L}$. 
    \item For $L \notin \mathcal{L}$, the most negative zero is given by $k = k_{18, L^-}$.
\end{itemize}

Imposing graded unitarity constraints is now almost immediate. First note that determinant-based graded unitarity constraints at $L \geqslant 6$ and $L \in \mathcal{L}$ rule out the region $k \in (k_{18, L}, k_{18, L^-})$. On the other hand, at $L \geqslant 6$ and $L \notin \mathcal{L}$, there are no new constraints from graded unitarity. When combined with the constraints obtained from a bespoke low-level analysis in \eqref{eq:e7lowlevelconst}, one obtains that all-level determinant-based graded unitarity constraints for $V_k(\mathfrak{e}_7)$ rule out all values of $k$, except the following.
\begin{equation} \label{eq:e7alllevel}
    k = \begin{cases}
        \textcolor{knownSCFTtext}{k_{18,q} = -18 + \frac{18}{q}: (18,q) = 1} \\
        \textcolor{conjecturedSCFTtext}{k_{2,q} = -18 + \frac{2}{q} \ \ : (2,q) = 1} \\
        \textcolor{knownSCFTtext}{-4} \\
         -8 \\
          -11
    \end{cases}
\end{equation}
The $V_k(\mathfrak{e}_7)$ affine VOAs with $k$ being the set of boundary-admissible levels above, are precisely the associated VOAs of ``class-I'' $D_q^{18}(\mathfrak{e}_7)$ Argyres--Douglas SCFTs under the SCFT/VOA correspondence. Furthermore, $V_{-4}(\mathfrak{e}_7)$ is the associated VOA of the rank-one $E_7$ Minahan--Nemeschansky SCFT \cite{Minahan:1996cj}. 

The affine VOAs corresponding to the levels $k = k_{2,q}$ are non-admissible and conjectured to be associated with the $D_{9q}^{18}(\mathfrak{e}_7)$. As with the non-admissible cases investigated earlier, it could be that the associated VOAs of these theories are instead non-trivial finite extensions of $V_{k_{2,q}}(\mathfrak{e}_7)$. On the other hand, the affine VOAs corresponding to the remaining levels are not currently known to be associated with any unitary 4d $\mathcal{N} = 2$ SCFTs. In either case, we leave open the possibility that a finer, state-based graded unitarity analysis may rule out some or all of these non-admissible levels. 

%%%%%%%%%
\subsection{$V_k(\mathfrak{e}_8)$}
%%%%%%%%%
In this case, the regularity criterion \eqref{eq:ADEregularity} leads to the following constraints on $A_{(t,\alpha)}$,
\begin{equation} \label{eq:Atconste8}
    A_{(t,\alpha)} \ \ \begin{cases}
        = 30 \qquad \text{when} \ \ t = 1 \ \ \text{and} \ \ \alpha = \theta \\ = 30 \qquad \text{when} \ \ t = 2 \ \ \text{and} \ \ \alpha = (1,1,2,3,2,2,1,1) =: \alpha^{\mathfrak{e}_8}_{(2)} \\ = 30 \qquad \text{when} \ \ t = 3 \ \ \text{and} \ \ \alpha = (0,1,1,2,1,1,1,0) =: \alpha^{\mathfrak{e}_8}_{(3)} \\ = 30 \qquad \text{when} \ \ t = 5 \ \ \text{and} \ \ \alpha = (0,0,0,0,1,0,0,0) =: \alpha^{\mathfrak{e}_8}_{(5)} \\ \geqslant 48 \qquad \text{otherwise}
    \end{cases}
\end{equation}
%%%%%%%%%
\input{E8Dynkin}
%%%%%%%%%

As in the previous cases, the regular dominant weight in the Weyl orbit of $2\alpha^{\mathfrak{e}_8}_{(2)}$ and $3\alpha^{\mathfrak{e}_8}_{(3)}$ is the highest root, leading to, 
\begin{equation}
    \frac{E(\theta)}{E(0)} = -\frac{E(2\alpha^{\mathfrak{e}_8}_{(2)})}{E(0)} = -\frac{E(3\alpha^{\mathfrak{e}_8}_{(3)})}{E(0)} = -\frac{E(5\alpha^{\mathfrak{e}_8}_{(5)})}{E(0)} = \text{ch Ad}~.
\end{equation}

For $V_k(\mathfrak{e}_8)$, note that the vacuum determinant in an appropriate fixed-weight sector for conformal levels $L \leqslant 7$ is given by, 
\begin{equation} \label{eq:lowlevele8det1to7}
\begin{split}
    &\text{det}~S_{\delta + \alpha} \ \ \ \propto \textcolor{blue}{(k - 0)}~, \\
    &\text{det}~S_{2\delta + \omega_1} \ \propto \textcolor{blue}{(k + 6)^1}(k - 0)^{\#} (\ldots \ldots)~, \\ 
    &\text{det}~S_{3\delta + \omega_7} \ \propto \textcolor{blue}{(k + 10)} (k + 6)^{\#} (\ldots \ldots)~, \\
    &\text{det}~S_{4\delta + 2\omega_8} \propto \textcolor{blue}{(k + 29/2)} (k + 10 )^{\#} (\ldots \ldots)~, \\
    &\text{det}~S_{5\delta + \omega_7} \ \propto \textcolor{blue}{(k + 18)} (k + 29/2)^{\#} (\ldots \ldots)~, \\
    &\text{det}~S_{6\delta + \omega_1} \ \propto \textcolor{blue}{(k + 22)^{1}(k + 59/3)^6}(k+18)^{\#} (\ldots \ldots)~, \\
    &\text{det}~S_{7\delta + \alpha} \ \ \propto \textcolor{blue}{(k + 180/7)^1}(k+22)^{\#} (\ldots \ldots)~.
\end{split}
\end{equation}
Determinant-based graded unitarity constraints from \eqref{eq:grunitsignrule} up to $L \leqslant 7$ and passing to a representation basis instead of a weight basis when necessary, we find that compatibility with graded unitarity restricts the level $k$ to be,
\begin{equation} \label{eq:e8lowlevelconst}
    k \leqslant -\frac{180}{7} \qquad \text{or} \qquad k = \left\{-22, -18, -\frac{29}{2}, -10, -6, 0 \right\}~.
\end{equation}

From the relevant data tabulated in Table \ref{tb:e8zerodata} and \eqref{eq:Atconste8}, we find that at conformal level $L$, the possible zeros in a fixed-root vacuum determinant are given by,
\begin{equation} \label{eq: e8prelimzero}
    k = \begin{cases}
        k_{30,q} = -30 + \frac{30}{q}: q \leqslant L \ \ \text{and} \ \ (q,30)=1 \\ 
        k_{2,q} = -30 + \frac{2}{q} \ \ : q \leqslant \lfloor \frac{L}{30} \rfloor \ \ \text{and} \ \ (q,2)=1 \\
        k_{3,q} = -30 + \frac{3}{q} \ \ : q \leqslant \lfloor \frac{L}{31} \rfloor \ \ \text{and} \ \ (q,3)=1 \\
        k_{5,q} = -30 + \frac{5}{q} \ \ : q \leqslant \lfloor \frac{L}{18} \rfloor \ \ \text{and} \ \ (q,5)=1 \\
        k_{6,q} = -30 + \frac{6}{q} \ \ : q \leqslant \lfloor \frac{L}{22} \rfloor \ \ \text{and} \ \ (q,6)=1 \\
        k_{10,q} = -30 + \frac{10}{q} : q \leqslant \lfloor \frac{L}{10} \rfloor \ \ \text{and} \ \ (q,10)=1 \\
        k_{15,q} = -30 + \frac{15}{q} : q \leqslant \lfloor \frac{L}{6} \rfloor \ \ \text{and} \ \ (q,15)=1 \\
        k_{A,B} = -30 + \frac{A}{B} : A \geqslant 48 \ \ \text{and} \ \ B \leqslant L
    \end{cases}
\end{equation}
%%%%%%%%%%%%%%%%%%%%%%%%%
\begin{table}[]
    \centering
    \begin{tabular}{|c|c|c|c|}
        \hline\hline
        $k$ & $L_s$ & Representation & Multiplicity \\
        \hline\hline
        $k_{2,q}$  & $30q$ & $V_{\omega_7} = \textbf{30380}$ & $35$ \\
        $k_{3,q}$  & $31q$  & $V_{\omega_7+\omega_8} = \textbf{4096000}$ & $1624$ \\
        $k_{5,q}$  & $18q$ & $V_{\omega_6} = \textbf{2450240}$ & $1056$ \\
        $k_{6,q}$  & $22q$  & $V_{4\omega_8} = \textbf{79143000}$ & $11705$ \\
        $k_{10,q}$ & $10q$   & $V_{2\omega_1} = \textbf{4881384}$ & $1584$ \\
        $k_{15,q}$ & $6q$   & $V_{\omega_6} = \textbf{2450240}$ & $1056$ \\
        $k_{30,q}$ & $q$   & $V_\theta = \textbf{248}$ & $1$ \\
        \hline
    \end{tabular}
    \caption{\label{tb:e8zerodata} Smallest conformal level $L_s$, primitive null representations and their root multiplicities for different values of $k$.}
\end{table}
%%%%%%%%%%%%%%%%%%%%%%%%

In addition, we note that the multiplicity of the $k = k_{30,q}$ zero in the conformal level $L = q$, fixed-root vacuum determinant is one, as the null state at this zero is a component of the adjoint representation. Motivated by the structure of possible zeros in \eqref{eq: e8prelimzero}, let us define the set $\mathcal{L} = \{1,7,11,13,17,19,23,29\}~\text{mod}~30$, comprising all positive integers coprime to thirty. Along with \eqref{eq: e8prelimzero} and the multiplicities documented in Table \ref{tb:e8zerodata}, this leads to the following structure of zeros for the level $L \geqslant 8$ fixed-root vacuum determinant.
\begin{itemize}
    \item For $L \in \mathcal{L}$, the most negative zero is given by $k = k_{30,L}$ and this zero always appears with multiplicity one. Furthermore, the second most negative zero is given by $k = k_{30,L^-}$. 
    \item For $L \notin \mathcal{L}$, the most negative zero is given by $k = k_{30, L^-}$.
\end{itemize}

The graded unitarity constraints now follow straightforwardly. Determinant-based graded unitarity constraints at $L \geqslant 8$ and $L \in \mathcal{L}$ rule out the region $k \in (k_{30, L}, k_{30, L^-})$. On the other hand, at $L \geqslant 8$ and $L \notin \mathcal{L}$, there are no new constraints. When combined with the results from the bespoke low-level analysis in \eqref{eq:e8lowlevelconst}, one obtains that all-level determinant-based graded unitarity constraints for $V_k(\mathfrak{e}_8)$ rule out all values of $k$, except the following.
\begin{equation} \label{eq:e8alllevel}
    k = \begin{cases}
        \textcolor{knownSCFTtext}{k_{30,q} = -30 + \frac{30}{q}: (30,q) = 1} \\
        \textcolor{knownSCFTtext}{-6} \\
        \textcolor{conjecturedSCFTtext}{-10, -18, -22} \\
        -\frac{29}{2}
    \end{cases}
\end{equation}
The $V_k(\mathfrak{e}_8)$ affine VOAs for  $k$ being the set of boundary-admissible levels above are precisely the associated VOAs of $D_q^{30}(\mathfrak{e}_8)$ ``class I" Argyres--Douglas SCFTs. Furthermore, $V_{-6}(\mathfrak{e}_8)$ is the associated VOA of the rank-one $E_8$ Minahan--Nemeschansky SCFT \cite{Minahan:1996cj}. 

The affine VOAs corresponding to the levels $-10, -18$, and $-22$ are non-admissible, but conjectured to be associated with the $D_{1}^{20}(\mathfrak{e}_8), D_{2}^{24}(\mathfrak{e}_8)$, and $D_{3}^{24}(\mathfrak{e}_8)$ Argyres--Douglas SCFTs respectively. As discussed in Section \ref{sec:intro}, the non-admissibility of these levels does not preclude the associated VOAs of these theories from being non-trivial finite extensions of the corresponding affine Kac--Moody algebras. On the other hand, the affine VOA $V_{-\frac{29}{2}}$, while admissible and quasi-lisse \cite{Arakawa:2010ni, Arakawa:2016hkg}, is not currently known to be associated with any unitary 4d $\mathcal{N} = 2$ SCFTs. Nevertheless, some or all of these levels may be excluded by a more refined state-based analysis of graded unitarity.

%% file: E6Dynkin.tex
%------------------------------------------------
% E_6
%------------------------------------------------
\begin{figure} 
\centering
\begin{tikzpicture}[scale=0.9,
    every node/.style={circle, draw, fill=white, inner sep=2pt, minimum size=6mm}]
    
    % Nodes
    \node (1) at (0,0)   {$1$};
    \node (2) at (1.5,0) {$3$};
    \node (3) at (3,0)   {$4$};
    \node (4) at (4.5,0) {$5$};
    \node (5) at (6,0) {$6$};
    \node (6) at (3,1.2) {$2$};

    % Edges
    \draw (1)--(2)--(3)--(4)--(5);
    \draw (3)--(6);

\end{tikzpicture}
\caption{Dynkin diagram of $E_6$} \label{fig:E6Dynkindiagram}
\end{figure} 

%% file: E7Dynkin.tex
%------------------------------------------------
% E_7
%------------------------------------------------
\begin{figure} \label{fig:E7Dynkindiagram}
\centering
\begin{tikzpicture}[scale=0.9,
    every node/.style={circle, draw, fill=white, inner sep=2pt, minimum size=6mm}]
    
    % Nodes
    \node (1) at (0,0)   {$1$};
    \node (2) at (1.5,0) {$3$};
    \node (3) at (3,0)   {$4$};
    \node (4) at (4.5,0) {$5$};
    \node (5) at (6,0)   {$6$};
    \node (6) at (7.5,0) {$7$};
    \node (7) at (3,1.2) {$2$};

    % Edges
    \draw (1)--(2)--(3)--(4)--(5)--(6);
    \draw (3)--(7);
\end{tikzpicture}
\caption{Dynkin diagram of $E_7$}
\end{figure}

%% file: E8Dynkin.tex
%------------------------------------------------
% E_8
%------------------------------------------------
\begin{figure} \label{fig:E8Dynkindiagram}
\centering
\begin{tikzpicture}[scale=0.9,
    every node/.style={circle, draw, fill=white, inner sep=2pt, minimum size=6mm}]
    
    % Nodes
    \node (1) at (0,0)   {$1$};
    \node (2) at (1.5,0) {$3$};
    \node (3) at (3,0)   {$4$};
    \node (4) at (4.5,0) {$5$};
    \node (5) at (6,0)   {$6$};
    \node (6) at (7.5,0) {$7$};
    \node (7) at (9,0)   {$8$};
    \node (8) at (3,1.2) {$2$};

    % Edges
    \draw (1)--(2)--(3)--(4)--(5)--(6)--(7);
    \draw (3)--(8);
\end{tikzpicture}
\caption{Dynkin diagram of $E_8$}
\end{figure}

%% file: sections/appendix_VkSOodd.tex
\section{Details of graded unitarity constraints on $V_k(\sof_{2N})$ with $N$ odd} \label{app:VkSOodd}
In this appendix, we provide the details of the graded unitarity analysis for $V_k(\sof_{2N})$ with $N$ odd. 
\subsection{Low level constraints}
We begin with a bespoke analysis for conformal levels $L \leqslant 5$. To this end, note that the vacuum determinant in a fixed-root sector for $L = 1$ and $L = 2$ is given by,
\begin{equation} \label{eq: so2NoddL1to2}
\begin{split}
    &\text{det}~S_{\delta + \alpha} \ \propto \textcolor{blue}{(k - 0)}~, \\
    &\text{det}~S_{2\delta + \alpha} \propto \textcolor{blue}{(k - k_{N,1})}(k-k_{2N-2,1})^{\#} (\ldots \ldots)~. \\ 
\end{split}   
\end{equation}
Thus, up to $L = 2$, graded unitarity constraints require the level $k$ to be in the regime $-(2N-2) < k \leqslant -N+2$ or $k = 0$. The fixed-root vacuum determinant at $L = 3$ is given by,
\begin{equation} \label{eq: so2NoddL3}
\begin{split}
    &\text{det}~S_{3\delta + \alpha} \propto \textcolor{blue}{(k-k_{8,3})(k-k_{4,1})^3}(k-k_{5,1})^{\#}(\ldots\ldots) \qquad \text{when} \ \ N=5~, \\
    &\text{det}~S_{3\delta + \alpha} \propto \textcolor{blue}{(k - k_{2N-2,3})}(k-k_{N,1})^{\#_2} (\ldots \ldots) \qquad \qquad \ \text{when} \ \ N \geqslant 7~.\\ 
\end{split}   
\end{equation}
For $N = 5$, the primitive null states at $k = k_{8,3}$ and $k = k_{4,1}$ are components of the adjoint, and the $V_{\omega_4+\omega_5}$ representation. Hence, passing to a representation basis instead of a weight basis, graded unitarity constraints rule out $k \in (k_{8,3}, k_{5,1})$. For $N \geqslant 7$, graded unitarity constraints of \eqref{eq:grunitsignrule} rule out $k \in (k_{2N-2,3}, k_{N,1})$. 

At conformal level $L = 4$, the fixed-root vacuum determinant is given by,
\begin{equation} \label{eq: so2NoddL4}
\begin{split}
    &\text{det}~S_{4\delta + \alpha} \propto \textcolor{blue}{(k-k_{2,1})(k-k_{5,2})}(k-k_{8,3})^{\#}(\ldots\ldots) \qquad \text{when} \ \ N=5~, \\
    &\text{det}~S_{4\delta + \alpha} \propto \textcolor{blue}{(k - k_{N,2})}(k-k_{2N-2,3})^{\#_2} (\ldots \ldots) \qquad \qquad \text{when} \ \ N \geqslant 7~.\\ 
\end{split}   
\end{equation}
For $N = 5$, the primitive null states at $k = k_{2,1}$ and $k = k_{5,2}$ are components of the adjoint and the $V_{2\omega_1}$ representation, respectively. Determinant-based graded unitarity constraints of \eqref{eq:grunitsignrule} rule out $k \in (k_{2,1}, k_{5,2})$. Moreover, passing to a representation basis instead of a weight basis, graded unitarity constraints rule out the entire interval $k \in (k_{2,1}, k_{8,3})$. For $N \geqslant 7$, graded unitarity constraints of \eqref{eq:grunitsignrule} rule out $k \in (k_{N,2}, k_{2N-2,3})$. 

Finally, at $L = 5$, the fixed-root vacuum determinant is given by,
\begin{equation} \label{eq: so2NoddL5}
\begin{split}
    &\text{det}~S_{5\delta + \alpha} \propto \textcolor{blue}{(k - k_{8,5})}(k-k_{2,1})^{\#} (\ldots \ldots) \qquad \qquad \text{when} \ \ N=5~, \\    
    &\text{det}~S_{5\delta + \alpha} \propto \textcolor{blue}{(k - k_{2N-2,5})}(k-k_{N,2})^{\#} (\ldots \ldots) \qquad \text{when} \ \ N \geqslant 7.
\end{split}
\end{equation}
Thus, up to $L = 5$, graded unitarity constraints restrict $k$ to be in the regime,
\begin{equation} \label{eq:lowlevelsoOdd}
\begin{split}
    &-(2N-2) < k \leqslant -\frac{32}{5} \qquad \text{or} \qquad k \in \left\{-6, -\frac{16}{3}, -3, 0 \right\}\qquad \ \ \ \text{when} \ \ N=5~,  \\
    &-(2N-2) < k \leqslant k_{2N-2,5} \ \ \ \text{or} \ \ \ \  k \in \{k_{2N,4}, k_{2N-2,3}, k_{2N,2}, 0\}  \ \ \ \text{when} \ \ N \geqslant 7~. 
\end{split}
\end{equation}

\subsection{Structure of zeros and constraints on the level}
With the possible zeros of the level $L$ vacuum determinant for $N$ odd given by \eqref{eq: so2nprelimzero}, we can infer the precise structure of zeros for the level $L$ vacuum determinant. For $N \geqslant 9$ and $6 \leqslant L \leqslant N-2$ (the $N = 5,7$ cases in this regime are handled by the low level analysis above), the structure of zeros follows from \eqref{eq:soevenstr1} and is identical to that in Section \ref{sec:soevenstruczero}. As with the case of $V_k(\mathfrak{so}_{2N})$ with $N$ even, determining the structure of zeros requires additional care, owing to the qualitatively different behaviour for $L = (N-1,N)~\text{mod}~(2N-2)$ compared to other conformal levels. 

To this end, first note that for $N = 5$, the structure of zeros up to $L = N = 5$ has already been documented in the low level analysis, and graded unitarity constraints up to $L = 5$ are summarised in \eqref{eq:lowlevelsoOdd}. On the other hand, for $N \geqslant 7$, determinant-based constraints from graded unitarity up to $L = N-2$ restrict the level $k$ to be in the regime given by,
\begin{equation} \label{eq:constSOodduptoNm2}
\scalebox{0.93}{$\displaystyle
    -(2N-2) < k \leqslant k_{2N-2, N-2} \ \ \text{or} \ \ k \in \{k_{2N, N-3}, k_{2N-2, N-4}, \ldots, k_{2N-2, 3}, k_{2N, 2}, k_{2N-2,1}\}~.
$}
\end{equation}
Furthermore, at $L = N-1$ and $L = N$, the structure of zeros takes the following form.
\begin{itemize}
    \item At $L = N-1$, the most negative zero of the fixed-root vacuum determinant is given by $k = k_{2N-2, N-1} = k_{2,1}$, and this zero always occurs with multiplicity one. The second most negative zero is $k = k_{2N,N-1}$ and this zero also occurs with multiplicity one.\footnote{Note that from \eqref{eq:Lmultofk2N}, $k = k_{2N,N-1}$ first appears as a zero at $L = N-1$, and the corresponding primitive null state is a component of the $V_{2\omega_1}$ representation. Since the root multiplicity of $V_{2\omega_1}$ is one, the multiplicity of this zero in the fixed-root, conformal level $L=N-1$ vacuum determinant is one.} In addition, the third most negative zero is given by $k = k_{2N-2, N-2}$.  
    \item At $L = N$, the most negative zero of the fixed-root vacuum determinant is given by $k = k_{2N-2,N}$, and this zero always occurs with multiplicity one. Furthermore, the second most negative zero is given by $k = k_{2, 1}$.  
\end{itemize}
The novel difference between the analysis for $N$ even and $N$ odd, first appears at $L = N$, and can be traced back to the contrasting multiplicities in \eqref{eq:Lmultofk2Nm2}. The above structure of zeros, together with the low-level analysis in \eqref{eq:lowlevelsoOdd}, immediately implies that determinant-based graded unitarity constraints up to $L = N$ rule out all levels $k$ except,\footnote{At $L = N-1$, determinant-based constraints from \eqref{eq:grunitsignrule} only eliminate $k \in (k_{2,1}, k_{2N, N-1})$. However, passing to a representation basis rather than a weight basis, since the primitive null representations at $k=k_{2,1}$ and $k = k_{2N,N-1}$ are different, the entire region $k \in (k_{2,1}, k_{2N-2, N-2})$ is ruled out.}
\begin{equation} \label{eq:so4np2levellessNp1}
    -(2N-2) < k \leqslant k_{2N-2,N} \ \ \text{or} \ \ k \in \{k_{2,1},k_{2N-2,i}: i \in \mathcal{I}^{\rm odd}_N, k_{2N,j}: j \in \mathcal{J}^{\rm odd}_N \}~.
\end{equation}
where $\mathcal{I}^{\rm odd}_N$ and $\mathcal{J}^{\rm odd}_N$ are the sets,
\begin{equation}
\begin{split}
    &\mathcal{I}^{\rm odd}_N = \{1, 3, 5, \ldots, N-2\}~, \\ 
    &\mathcal{J}^{\rm odd}_N = \{2, 4, 6, \ldots, N-3\}~.
\end{split}
\end{equation}

For $N \geqslant 5$ and $L \geqslant N+1$, it can be shown by a direct computation that the fixed-root vacuum determinant takes the following form, \\
\\
\underline{\textbf{Case 1: } $L \neq (N-1,N)~\text{mod}~(2N-2)$}
\begin{equation}
\begin{split}
    &\text{det}~S_{L\delta+\alpha} \propto \textcolor{blue}{(k-k_{2N-2,L})^1}(k-k_{2N-2,L-2})^{\#}( \ldots \ldots) \qquad L \ \ \text{odd}~, \\
    &\text{det}~S_{L\delta+\alpha} \propto \textcolor{blue}{(k-k_{2N-2,L-1})^{\#}}( \ldots \ldots) \qquad \qquad \qquad \qquad L \ \ \text{even}~.
\end{split}
\end{equation}
\underline{\textbf{Case 2: } $L = (N-1,N)~\text{mod}~(2N-2)$}
\begin{equation}
\begin{split}
    &\text{det}~S_{L\delta+\alpha} \propto \textcolor{blue}{(k-k_{2N-2,L})^1}(k-k_{2N-2,L-1})^{\#}( \ldots \ldots)~.
\end{split}
\end{equation}
With this structure of zeros of the vacuum determinant, all-level determinant-based graded unitarity constraints are straightforward to implement. Note that for $L \neq (N-1,N)~\text{mod}~(2N-2)$, graded unitarity sign constraints of \eqref{eq:grunitsignrule} eliminate the region $k \in (k_{2N-2, L}, k_{2N-2, L-2})$ for $L$ odd, whereas, it does not give any constraints for $L$ even. Whereas for $L = (N-1,N)~\text{mod}~(2N-2)$, graded unitarity sign constraints rule out $k \in (k_{2N-2, L}, k_{2N-2, L-1})$. Combining this with the graded unitarity constraints up to $L = N$, \emph{i.e.}, \eqref{eq:so4np2levellessNp1}, we find that under the assumption of the standard $\mfR$-filtration \eqref{eq:AKMRfiltration} on $V_k(\mathfrak{so}_{2N})$ affine VOAs, fixed-root determinant-based graded unitarity rules out all levels $k$ except,
\begin{equation} \label{eq:VkSOoddgrunitresults}
k = \begin{cases}
        -(2N-2) + \frac{(2N-2)}{q}: (2N-2, q) = 1 \\
        -(2N-2) + \frac{(2N-2)}{q}: q \in 2\mathbb{N}+1 \ \ \text{and} \ \ (2N-2, q) \neq 1 \\
        -(2N-2) + \frac{N}{q} \qquad : q \in \mathbb{N} \ \ \text{and} \ \ 1 \leqslant q \leqslant \frac{N-3}{2} \\
        -(2N-2) + \frac{2}{q} \qquad \ : q \in 2\mathbb{N}+1 \\     
    \end{cases}
\end{equation}
Combining these graded unitarity restrictions on the level $k$ with those for $V_k(\mathfrak{so}_{2N})$ with $N$ even, we get the final result \eqref{eq:VkSOfullresults}.